\documentclass[10pt, conference]{IEEEtran}
\usepackage[colorlinks]{hyperref}
\usepackage{amssymb}
\usepackage{algorithm}
\usepackage{algorithmic}
\usepackage{xcolor}
\usepackage{mathrsfs}
\usepackage{blkarray}
\usepackage{amsthm}

\ifCLASSINFOpdf
  \usepackage[pdftex]{graphicx}
  \graphicspath{{./figures/}}
  \DeclareGraphicsExtensions{.pdf,.jpeg,.png}
\else
\fi

\usepackage{amsfonts}
\usepackage{amsmath}
\usepackage{caption}
\usepackage{subcaption}
\usepackage[inline]{enumitem}

\usepackage{algorithm}
\usepackage{algorithmic}
\usepackage{xcolor}
\usepackage{mathrsfs}
\usepackage{blkarray}
\usepackage{amsthm}

\usepackage{circledsteps}

\ifCLASSINFOpdf
\else
\fi
\usepackage{amsmath}
\usepackage{amsfonts}
\begin{document}
%
\title{Model-Based Deep Learning Tuning of Reconfigurable Intelligent Surface for OFDM Radar Interference Mitigation}
%
%
%

\author{\IEEEauthorblockN{{Ali Parchekani}, {Milad Johnny}, and {Shahrokh Valaee} }
\IEEEauthorblockA{Department of Electrical and Computer Engineering, \\
University of Toronto, Canada \\
 Email: a.parchekani@mail.utoronto.ca, milad.johnny@utoronto.ca, and valaee@ece.utoronto.ca}}

\maketitle

\begin{abstract}
This paper presents a deep learning-based framework for enhancing radar systems in the presence of interference, leveraging Reconfigurable Intelligent Surfaces (RIS). The proposed technique uses a modified MUSIC algorithm to estimate the angles of the target and interference. The core of the method is a deep learning model that optimizes the RIS configuration to reduce the impact of interference while maintaining accurate angle estimates. The model consists of a multi-layer perceptron (MLP) that takes estimated angles as inputs and outputs the configuration of the RIS. A specially designed loss function ensures that the interference is properly suppressed and the target remains detectable. To further enhance performance, a convolution technique is introduced to create a notch at the interference angle, ensuring better separation between the target and interference. Additionally, the method is extended to work over multiple subcarriers, improving robustness and performance in practical scenarios. Simulation results show that the technique enhances the signal-to-interference-plus-noise ratio (SINR) and provides accurate localization estimates, demonstrating its potential for radar systems in complex environments.
\end{abstract}


%
\IEEEpeerreviewmaketitle

\section{Introduction}

Orthogonal Frequency Division Multiplexing (OFDM) radar has emerged as a prominent waveform for modern radar systems \cite{ofdmradar}, particularly in automotive and integrated sensing and communication applications \cite{ofdmsensing}. By transmitting multiple orthogonal subcarriers, OFDM radars can jointly perform high-resolution sensing and data communication on shared spectrum \cite{jointsensing}. In automotive scenarios, OFDM radars enable vehicles to detect obstacles and measure range-velocity \cite{automotiveradar}. A critical challenge faced by these radars is mutual interference. When multiple OFDM radars operate in proximity, their signals can interfere with each other and drastically degrade detection performance \cite{chalmers}. Effective interference mitigation is crucial to preserve the reliability and accuracy of OFDM radar sensing in practical deployments.

Various techniques have been explored to combat radar interference. Traditional approaches include temporal or frequency multiplexing of radar signals to avoid overlapping transmissions, but such coordination is not always feasible in automotive scenarios. Signal processing methods have been proposed to suppress interference, such as compressed sensing techniques to recover target signal from interfered OFDM signal \cite{compressedsensing}. A cooperative approach was introduced in the form of repeated-symbol OFDM (RS-OFDM), where radars intentionally align symbols to enable interference cancellation \cite{rsofdm}. While effective under specific conditions, many of these solutions require specialized waveforms, coordination between radars, or complex signal reconstruction, which may limit their practicality. There is a growing need for more flexible interference mitigation strategies that do not impose stringent requirements on radar transmitter and can adapt to dynamic interference conditions.

Reconfigurable Intelligent Surface (RIS) has recently gained attention as an emerging technology capable of controlling the wireless propagation environment \cite{RISwireless}. RIS consists of array of passive scattering elements that can modify the phase and amplitude of the incident signal \cite{phasechange}. By appropriately configuring the phase shifts across the RIS, the reflection of the signals can be customized to steer energy in desired directions or even cancel out signals from undesired directions \cite{convpaper}. RIS has been applied in Integrated Sensing and Communication (ISAC) systems \cite{mustafamag}, where it has been used to balance the dual needs of radar and communication on the same spectrum \cite{dualradar}. It has also been used to improve positioning \cite{mustafanew} and improve energy efficiency \cite{javadkalbasiglobecom}.
RIS can be used for interference mitigation in radar domain. Instead of treating the propagation channel as fixed, RIS allows the environment to be modified in real-time to favor radar's operation. It can be used to improve the gain of the target's path while reducing the effect of the interference radar \cite{convpaper}. This can be done without modification of the radar waveforms and does not require cooperation between them. While interference mitigation clears the way for detecting the targets, the radar must also accurately estimate the angles or directions-of-arrival (DoA) of the target echoes. In automotive and wireless sensing, knowing the angle of the object is as important as estimating range and velocity. Furthermore, RIS operation depends on accurate power reduction of the interference. 

Model based deep learning \cite{modelbased} is a promising approach that combines the rigor of the established models with the flexibility of deep neural networks. It uses the domain-specific knowledge with the structure of the network to improve various aspects of the problem such as parameter tuning while maintaining the theoretical foundations. 

In this work, we utilize RIS to perform both angle estimation as well as interference mitigation. We use the Music algorithm to estimate the directions of the interference and target signals and use RIS to reduce the effect of interference and reinforce the power of the target. We propose to use a model-based deep learning approach to tune the elements of the RIS for both estimation and beam pattern design. We show the effectiveness of the proposed method and its high resolution in detection of targets and interference, while reducing the effect of the interfering radar. 

The remainder of the paper is organized as follows. Next section describes the system model. Section \ref{sec:problem} describes the problem and the proposed method to address the problem. Section \ref{sec:simulation} represents the simulation results of the problem, and Section \ref{sec:conclusion} concludes the paper. 

\section{System Model} \label{sec:model}

\begin{figure}
    \centering
    \includegraphics[width=0.8\linewidth]{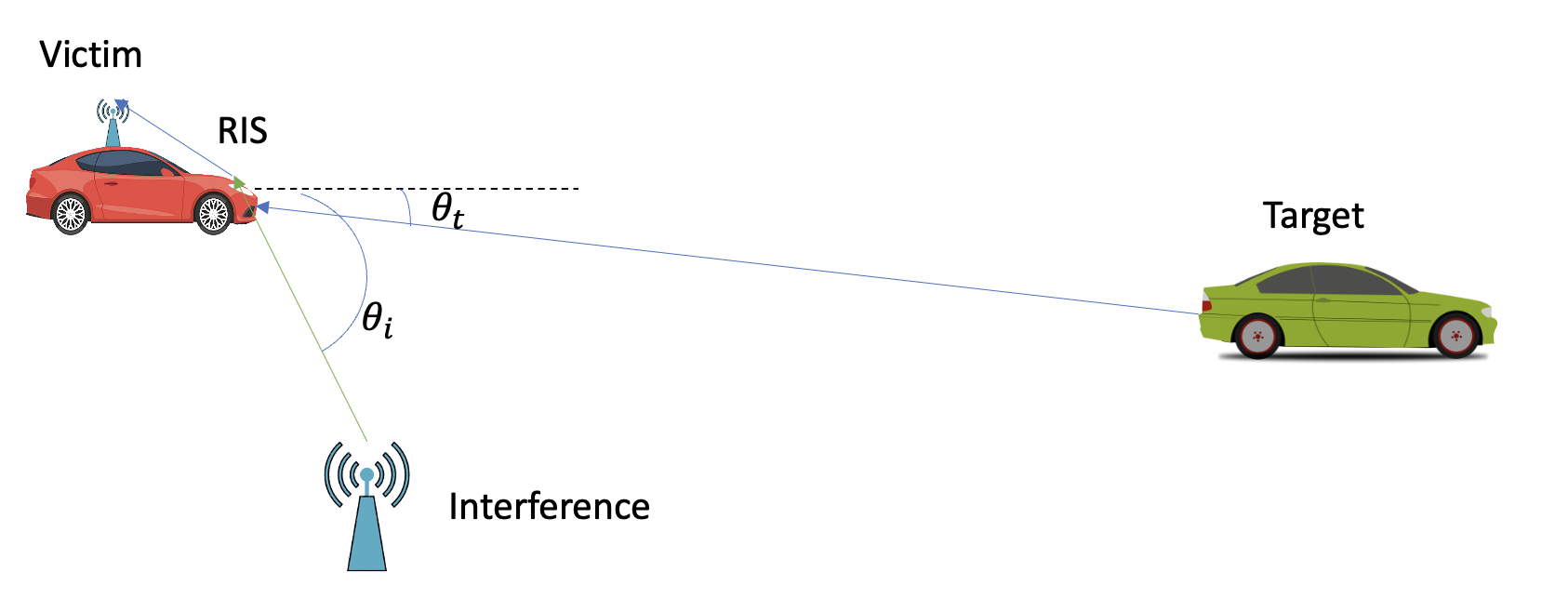}
    \caption{Setup of the problem}
    \label{fig:scenario}
\end{figure}

Consider the scenario illustrated in Fig.~\ref{fig:scenario}. It involves a car of interest equipped with a single transceiver antenna mounted on its roof, while its hood is coated with a RIS. The scenario begins with the antenna transmitting an OFDM signal into the environment. Additionally, a target is present in the environment, along with another radar that acts as an interference for the car of interest. The transmitted OFDM signal in a transmission block of duration $T$ contains $N$ subcarriers, each carrying $M$ symbols. Therefore, it can be expressed as:
 \begin{equation}
 \label{original1}
 s(t) = \sum_{n = 0}^{N -1} \sum_{m = 0}^{M -1} d^{(v)}(n,m) e^{j 2 \pi (f_c + n \Delta f) t} \text{rect}\left( \frac{t - mT}{T}\right), \end{equation}
where $d^{(v)}(n,m)$ is the transmitted symbol chosen from an arbitrary modulation scheme such as the phase-shift keying (PSK) at time $m$ over the subcarrier $n$, $f_c$ is the carrier frequency, $\Delta f $ is the subcarrier spacing, and $T = T_{\mathrm{sym}} + T_{\mathrm{cp}}$ with $T_{\mathrm{sym}} = 1 / \Delta f$ and $T_{\mathrm{cp}}$ is the duration of the cyclic prefix to mitigate the effect of intersymbol interference (ISI).
For simplicity, the signal can be written as:
\begin{equation}
\label{originaleq}
   s(t) = \sum_{n = 0}^{N - 1} \sum_{m =  0}^{M - 1} s_{n,m}(t).
\end{equation}

The transmitted signal hits the target located at angle $\theta_t$ with the relative distance of $R_t$ and the relative velocity of $v_t$. This signal is reflected back to the RIS on the hood of the car. It is assumed that RIS has $L$ elements with the spacing of $\lambda/2$ where $\lambda = c / f_c$ with $c$ as the speed of wave propagation in the medium. The received signal reflected from the elements of the RIS can be written as:
\begin{align}
   s_{r, \mathrm{RIS}}(t) &= \sum_{m = 0}^{M - 1} \sum_{n = 0}^{N -1} \sum_{l = 1}^{L }   c_{m,l} e^{-j 2\pi \frac{d_l}{\lambda_n}} \\ \nonumber & \times  e^{-j2\pi \frac{\lambda}{2\lambda_n} (l-1) \sin(\theta_t)}  s_{n,m}(t-\tau + t \nu),
\end{align}
where $c_{m,l} \in \mathbb{C}$ represents the configuration of the $l$-th RIS element at time $m$, $d_l$ is the distance between the RIS element and the receiver antenna, $\lambda_n = c / (f_c + n \Delta f)$ is the subcarrier wavelength, $\tau = \frac{2R_t}{c}$ is the delay of the target, and $\nu = \frac{2v_t}{c}$ is the normalized Doppler shift of the target.

Let us define the array steering vector $\boldsymbol{b}_n(\theta) = [e^{-j \frac{2\pi d_1}{\lambda_n}}, \ldots, e^{-j \frac{2\pi d_L}{\lambda_n}}e^{-j \frac{2\pi \lambda}{2\lambda_n} (L - 1) \sin(\theta)}]^T \in \mathbb{C}^{L \times 1}$. By considering the OFDM signal received at various timeslots of the OFDM frame, it is possible to combine the configurations of the RIS elements over different timeslots in a matrix as:
\begin{equation}
    \boldsymbol{C} = \begin{bmatrix} 
c_{01} & c_{11} & \ldots & c_{(M - 1)1} \\
c_{02} & c_{12} & \ldots & c_{(M-1)2} \\
\vdots & \vdots & \ddots & \vdots \\
c_{0L} & c_{1L} & \ldots & c_{(M-1)L}
\end{bmatrix} ,
\end{equation}
where column $m \in \{0, \ldots, M - 1\}$ represents the configurations of the elements of the RIS at the $m$-th symbol time in a block with the time duration of $T$.
By defining $ \boldsymbol{S}_n = [s_{n, 0}(t - \tau + t \nu) , s_{n, 1}(t - \tau + t \nu), \ldots, s_{n, M - 1}(t - \tau + t \nu)]^T$, the received signal from angle $\theta$ can be written as:
\begin{equation} \label{target_ris}
    s_{r, RIS}(t) = \sum_{n = 0}^{N -1}  \boldsymbol{S}_n^T \boldsymbol{C}^T \boldsymbol{b}_n(\theta).
\end{equation}

Consider another OFDM radar operating near the radar of interest. Similar to \eqref{original1}, consider the transmitted signal of the interfering radar as:
\begin{align}
    s_I(t) &= \sum_{m = 0}^{M -1} \sum_{n = 0}^{N -1} d^{(i)}(n,m) e^{j 2 \pi (f_c + n \Delta f) t} \text{rect}\left( \frac{t - mT}{T}\right),
\end{align}
where $d^{(i)}(n,m)$ shows the transmitted symbols of the interfering radar. It is possible to rewrite the interference radar's signal as:
\begin{equation}
    s_I(t) = \sum_{m = 0}^{M -1} \sum_{n = 0}^{N -1} s_{I,n,m}(t).
\end{equation}

Let us assume that the interfering radar has the angle $\theta_i$ from the victim radar and has delay and normalized Doppler shift of $\tau_i$ and $\nu_i$, respectively. This signal hits the RIS and is received at the receiver antenna. The received signal would be expressed as:
\begin{align}
    s_{r, I}(t) = \sum_{m = 0}^{M - 1} \sum_{n = 0}^{N - 1} \sum_{l = 0}^{L -1}   c_{m,l} e^{-j 2\pi \frac{d_l}{\lambda_n}} \\ \nonumber \times e^{-j2\pi \frac{\lambda}{2\lambda_n}l  \sin(\theta_i)} s_{I,n,m}(t - \tau_i + t \nu_i)
\end{align}
By using the matrix notation and defining $\boldsymbol{S_{I,n}} = [s_{I,n, 0}(t - \tau + t \nu) , s_{I,n, 1}(t - \tau + t \nu), \ldots, s_{I,n, M - 1}(t - \tau + t \nu)]^T$:
\begin{equation} \label{interference_ris}
    s_{I,r}(t) = \sum_{n = 0}^{N -1}  \boldsymbol{S_{I,n}}^T  \boldsymbol{C}^T \boldsymbol{b}_n(\theta_i) 
\end{equation}
The total received signal at the receiver antenna can be expressed as:
\begin{equation}
    s_{r}(t) = s_{LoS}(t) + s_{r, \mathrm{RIS}}(t) + s_{I,r}(t) + n(t)
\end{equation} 
where $s_{LoS}(t)$ is the signal received through line-of-sight and $n(t)$ represents the additive noise of the environment with power $\sigma^2$. Assuming a stationary scenario and considering the configurations of the RIS elements as $\boldsymbol{C}_{t}$ and $\boldsymbol{C}_{t + 1}$ at two consecutive time slots with the same transmission symbol vector, we establish the relation $\boldsymbol{C}_t = -\boldsymbol{C}_{t + 1}$. Subtracting the two consecutive time signals allows the removal of the energy of the line-of-sight paths while preserving the energy of the paths from the RIS to the receiver.
Hence, it is assumed that the above procedure is done over the received signals.

The following steps are performed to identify the location and velocity of the target. First, the cyclic prefix is removed, then Fourier transform over the duration of the signal is applied, and the result is divided element-wise by the transmitted symbols. Then, for the $n$-th subcarrier of the $m$-th OFDM symbol, we have:
\begin{align}
    y[n,m] &= \alpha e^{-j 2\pi n \Delta f \tau} e^{j 2\pi f_c \nu m T}  \\ \nonumber &+ \alpha_i \frac{d^{(i)}(n,m)}{d^{(v)}(n,m)} 
    e^{-j 2\pi n \Delta f \tau_i} e^{j 2\pi f_c \nu_i m T} + z[n,m].
\end{align}
where, $z[n,m]$ represents the discrete noise term, $\alpha$ and $\alpha_i$ represent the gains of the RIS elements after the transformations. By taking the Fourier transform and the inverse Fourier transform over the $m$-th index and the $n$-th index of the first term respectively, we can find the delay and Doppler shift of the target. As it can be seen,  the symbols of the interference may cause error in estimating the target location. The goal is to use the configurations of the elements of the RIS in (\ref{target_ris}) and (\ref{interference_ris}) to increase the power of the target signal and reduce the effect of the interference.

\section{Problem Description \& Proposed Method} \label{sec:problem}

In this section, we describe the problem and propose a method to solve it. Based on the received signal, the angle and distance of both the target and the interfering radar, relative to the RIS and the receiver antenna, are unknown. The ultimate goal is to estimate the location of the target in the presence of interference. To achieve this, we employ Maximum Likelihood Estimation (MLE) by performing element-wise division of the received symbols by the transmitted symbols. We then apply the Fourier transform and the inverse Fourier transform over the subcarriers and time frame indices. The resulting term forms a range-Doppler map (RV-map). In the absence of interference, the term corresponding to the target simplifies to:
\begin{align} \label{target_rv}
    y[\hat{\tau}, \hat{\nu}] = \sum_{m = 0}^{M - 1} \sum_{n = 0}^{N - 1} \alpha  
    e^{-j 2\pi n \Delta f \tau_t}  e^{j 2\pi f_c \nu_i m T} \nonumber \\ e^{j 2 \pi n \Delta f \hat{\tau}} e^{-j 2 \pi m T f_c \hat{\nu}}.
\end{align}
The peaks of the above term would correspond to the range and Doppler shift of the target.
The issue with the presence of the interference is that the term corresponding to interference would have higher power compared to other terms and there is a mismatch between the interference symbols and the original radar's symbols causing the term corresponding to the interference radar to become:
\begin{align} \label{int_rv}
    y^{(i)}[\hat{\tau}, \hat{\nu}] &= \sum_{m = 0}^{M - 1} \sum_{n = 0}^{N - 1} \alpha_i \frac{d^{(i)}(n,m)}{d^{(v)}(n,m)} 
    e^{-j 2\pi n \Delta f \tau_i}  \\ \nonumber &\times e^{j 2\pi f_c \nu_i m T}  e^{j 2 \pi n \Delta f \hat{\tau}} e^{-j 2 \pi m T f_c \hat{\nu}}.
\end{align}
Depending on the power of the interfering radar, (\ref{int_rv}) can dominate the peaks of (\ref{target_rv}), exhibiting a noise-like behavior and thereby disrupting the estimation process.  

To address this issue, the RIS can be utilized to mitigate the effect of the interfering radar. Specifically, the beamforming capabilities of the RIS can be employed to block the direction of the interfering radar, which requires knowledge of the interference angle as well as proper tuning of the RIS elements. To achieve this, we divide the problem into two parts. First, we formulate a method to estimate the directions of both the interference and the target. Second, we design a beam pattern that suppresses the effect of the interfering radar while enhancing the target's direction. We propose using model-based deep learning for this purpose.  

The first part of the problem involves determining the directions of both the target and the interfering radar. Model-based algorithms such as MUSIC can be employed for this task. The MUSIC algorithm typically requires spatial diversity through multiple antennas to estimate the directions of arrival. However, in this setup, a single antenna is used for both transmission and reception. With the presence of the RIS, we propose modifying the algorithm to enable angle-of-arrival detection.  

\begin{figure}
    \centering
    \includegraphics[width=0.8\linewidth]{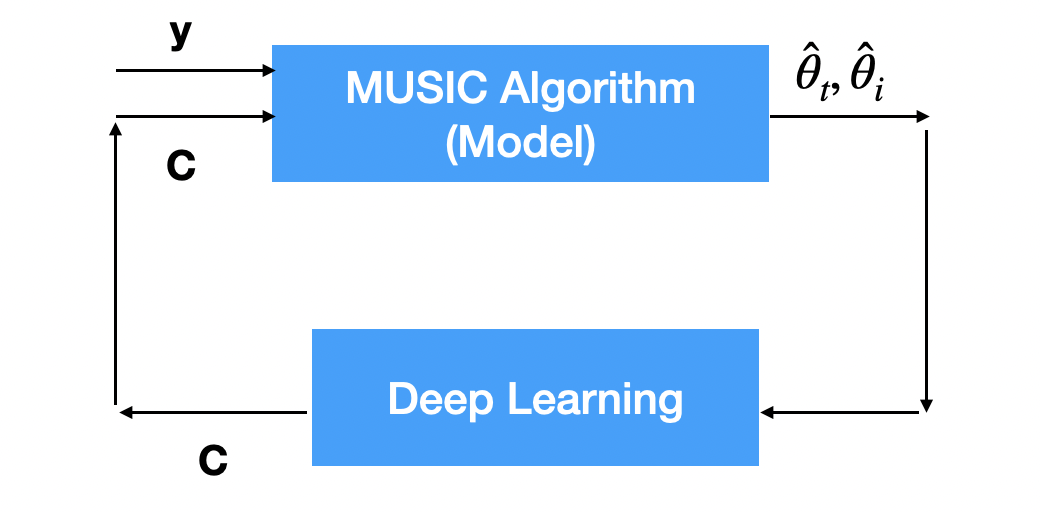}
    \caption{Model-based deep learning model structure}
    \label{fig:singlemodel}
\end{figure}

Consider the received signal. It contains the OFDM frame over different time slots with all subcarriers. Without loss of generality, let us consider one of the subcarriers, for instance the first one. Considering a stationary scenario, the signal of the subcarrier at different timeslots can be written as:
\begin{equation}
    \boldsymbol{y} = \eta_t \boldsymbol{C}^T \boldsymbol{b}(\theta_t) + \eta_i \boldsymbol{C}^T \boldsymbol{b}(\theta_i) + \boldsymbol{n},
\end{equation}
where $\boldsymbol{b(.)}$ represents the steering vector for the chosen subcarrier, $\eta_t$ and $\eta_i$ are the effective gains of the channel for the target and interference. Let us define $\boldsymbol{\phi}(\theta) = \boldsymbol{C}^T \boldsymbol{b}(\theta)$ as the modified steering vector. Hence, the received signal can be expressed as:
\begin{equation}
    \boldsymbol{y} = \boldsymbol{\Phi} \boldsymbol{\eta},
\end{equation}
where $\boldsymbol{\Phi} = [\boldsymbol{\phi}(\theta_t) \, \boldsymbol{\phi}(\theta_i)]$ and $\boldsymbol{\eta} = [\eta_t \, \eta_i]$.
The correlation matrix for the received signal can be written as:
\begin{equation}
    \boldsymbol{R} = E[\boldsymbol{\Phi}\boldsymbol{\eta} \boldsymbol{\eta}^H \boldsymbol{\Phi}^H] + E[\boldsymbol{n}\boldsymbol{n}^H]= \boldsymbol{\Phi} \boldsymbol{A}\boldsymbol{\Phi}^H + \sigma^2 \boldsymbol{I} = \boldsymbol{R}_s + \sigma^2 \boldsymbol{I},
\end{equation}
where $\boldsymbol{A} = E[\boldsymbol{\eta} \boldsymbol{\eta}^H]$ and $\boldsymbol{R}_s = \boldsymbol{\Phi} \boldsymbol{A}\boldsymbol{\Phi}^H$.
The above matrix has two non-zero eigenvalues corresponding to the angles of the target and the interference radars. The rest of the eigenvalues are zero. Consider one of the eigenvectors corresponding to the zero eigenvalues as $\boldsymbol{q}_m$, hence:
\begin{align} \label{eq:music}
    & \boldsymbol{R}_s\boldsymbol{q}_m = \boldsymbol{\Phi} \boldsymbol{A}\boldsymbol{\Phi}^H \boldsymbol{q}_m = 0 \\
    & \boldsymbol{q}_m^H\boldsymbol{\Phi} \boldsymbol{A}\boldsymbol{\Phi}^H \boldsymbol{q}_m = 0 \\ 
    & \boldsymbol{\Phi}^H \boldsymbol{q}_m = 0.
\end{align}

Equation (\ref{eq:music}) implies that all the eigenvectors of the zero eigenvalues are orthogonal to the modified steering vectors of the target and the interference radars. Hence, one can form a spectrum as:
\begin{equation} \label{eq:musiceq}
    P(\theta) =  \frac{1}{\lVert \boldsymbol{Q}_n^H \boldsymbol{C}^T \boldsymbol{b}(\theta) \rVert_2^2}
\end{equation}
where $\boldsymbol{Q}_n$ corresponds to all the eigenvectors of the zero eigenvalues. The peaks of \eqref{eq:musiceq} correspond to the angles of the target and the interference.

After determining the angles of the target and the interference radars, one should define a problem to reduce the effect of interference while increasing the power of the target signal. For that, it is possible to use the Signal-to-Interference-plus-Noise (SINR) ratio and tune RIS elements to maximize this ratio:
\begin{equation} \label{opteqn}
    \hat{\boldsymbol{C}} = \mbox{arg} \max_{\boldsymbol{C}}  \frac{ \lVert \boldsymbol{C}^T \boldsymbol{b}(\theta_t)\rVert_2^2}{\lVert \boldsymbol{C}^T \boldsymbol{b}(\theta_i)\rVert_2^2 + \sigma^2}.
\end{equation}

\begin{figure}
    \centering
    \includegraphics[width=0.8\linewidth]{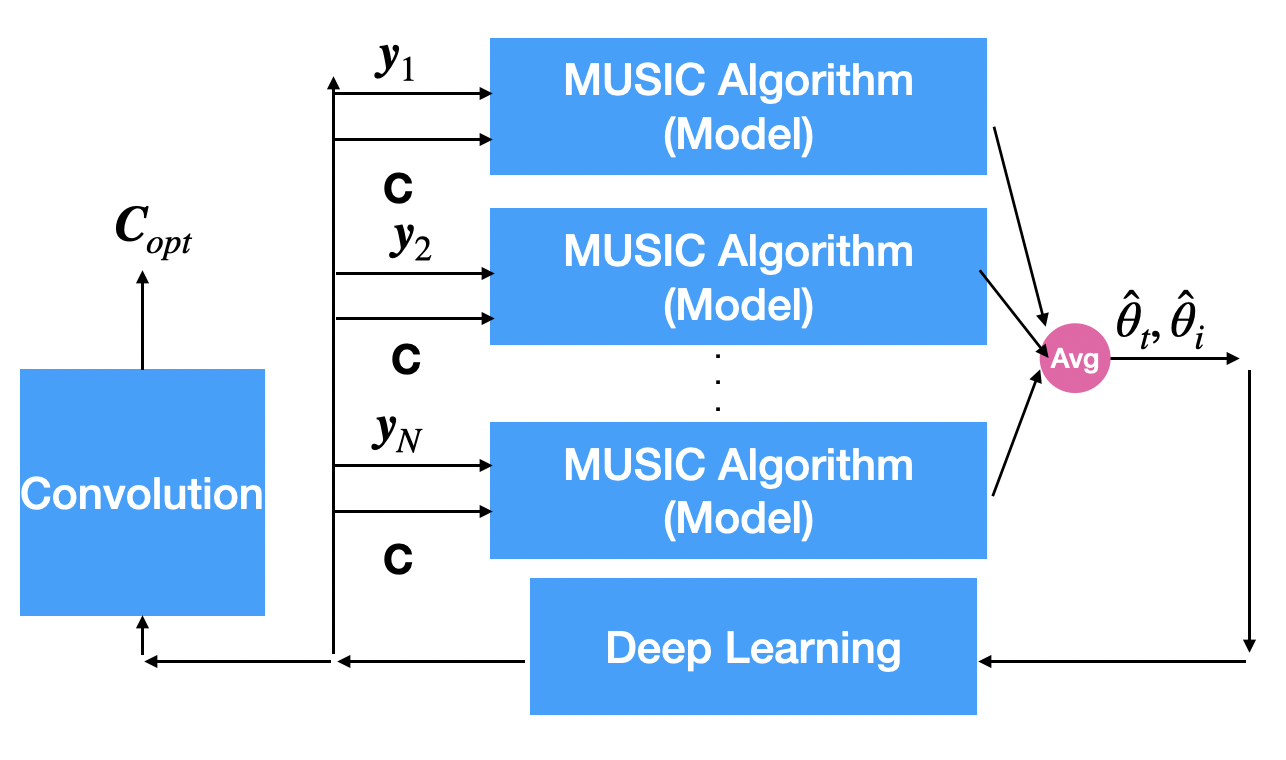}
    \caption{Structure of the model-based deep learning model.}
    \label{fig:structure}
\end{figure}

By exploring (\ref{opteqn}) and (\ref{eq:musiceq}), it can be seen that $\boldsymbol{C}$ acts as a tuning parameter in the former and as a hyperparameter in later. Hence, modifying it in the former would impact the latter. This motivates to use model-based deep learning to simultaneously tune RIS configurations for both angle estimation and SINR optimization. The idea of the model-based deep learning is to use deep learning alongside the model based algorithm to solve the problem. In this case, the parameter of interest is the configurations of the RIS elements and the modified MUSIC algorithm is the model. The proposed structure of the deep learning model is depicted in Fig.~\ref{fig:singlemodel}. The model consists of a loop in which the angles of target and the interference are estimated and the RIS configuration is tuned. The structure of the deep learning model can be Multi-Layer Perceptron (MLP) with the estimated angles of interference and target as the inputs and the configurations of RIS as the output. Hence, the loss function for this network plays a major role as it should tune the configuration to reduce the effect of interference while keeping the estimate of the angles consistent by maintaining the rank of the correlation matrix to be two and ensuring the peaks of both interference and target angle are detectable in the model. Therefore, the loss is expressed as:
\begin{equation}
    L = \beta \left[\frac{\lVert \boldsymbol{Q}_n^H \boldsymbol{C}^T \boldsymbol{b}(\hat{\theta_i}) \rVert_2^2}{\lVert \boldsymbol{Q}_n^H \boldsymbol{C}^T \boldsymbol{b}(\hat{\theta_t}) \rVert_2^2}\right] + (1-\beta ) \frac{\lVert \boldsymbol{C}^T \boldsymbol{b}(\hat{\theta_i})\rVert_2^2 + \sigma^2}{ \lVert \boldsymbol{C}^T \boldsymbol{b}(\hat{\theta_t})\rVert_2^2}.
\end{equation}
The proposed loss function has two parts, the first part ensures that the peak of the interference angle is stronger than the peak of the target angle in the modified MUSIC algorithm, this is to ensure that the rank of the correlation matrix is preserved and since the interference has stronger power than the target, stronger peak can be used to distinguish interference from the target. The second part of the loss is to increase the SINR by reducing the power received from the interference angle and increasing the power of the target angle.

\begin{figure*}[ht]
\centering
    \begin{subfigure}{.18\linewidth}
        \centering
        \includegraphics[width = 0.95 \linewidth]{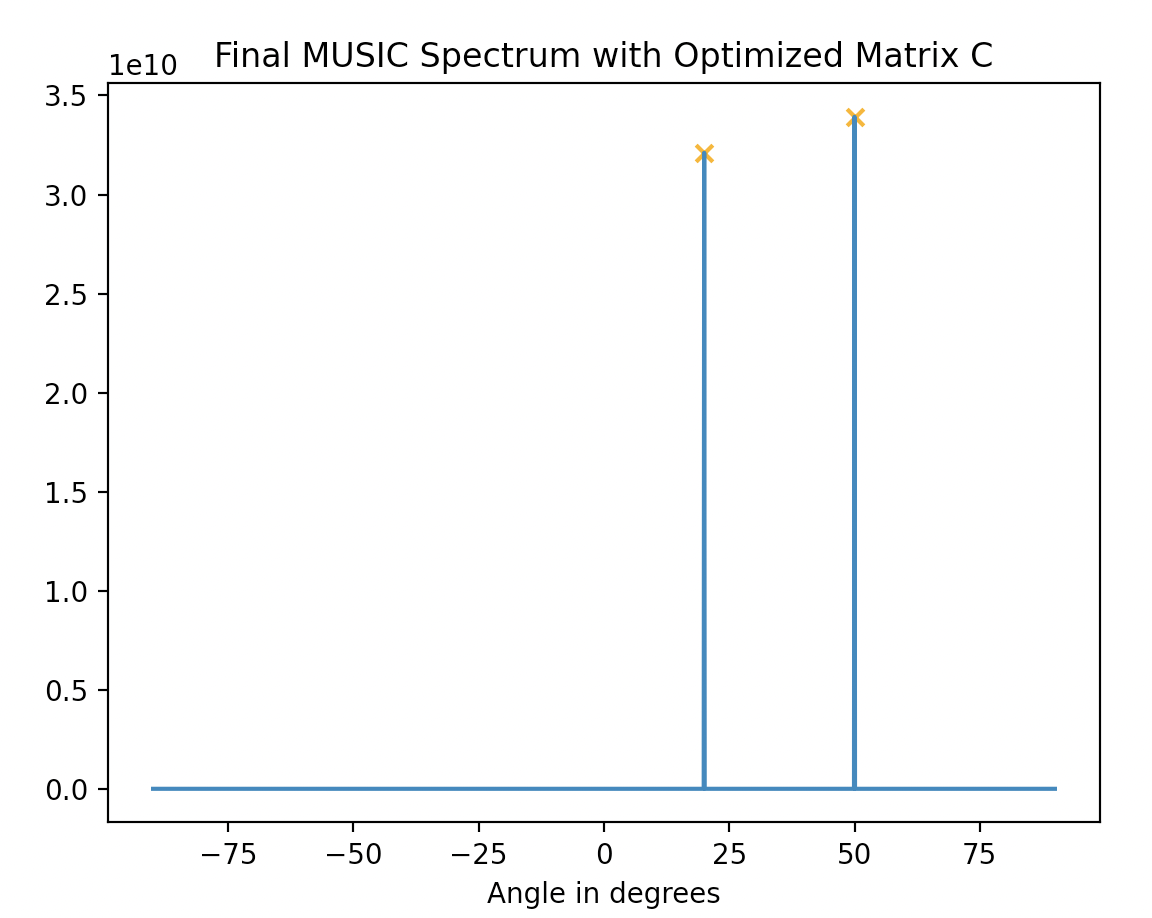}
        \caption{MUSIC spectrum for $\beta = 0$}
        \label{fig:0_spect}
    \end{subfigure}
    \begin{subfigure}{.18\linewidth}
        \centering
        \includegraphics[width = 0.95 \linewidth]{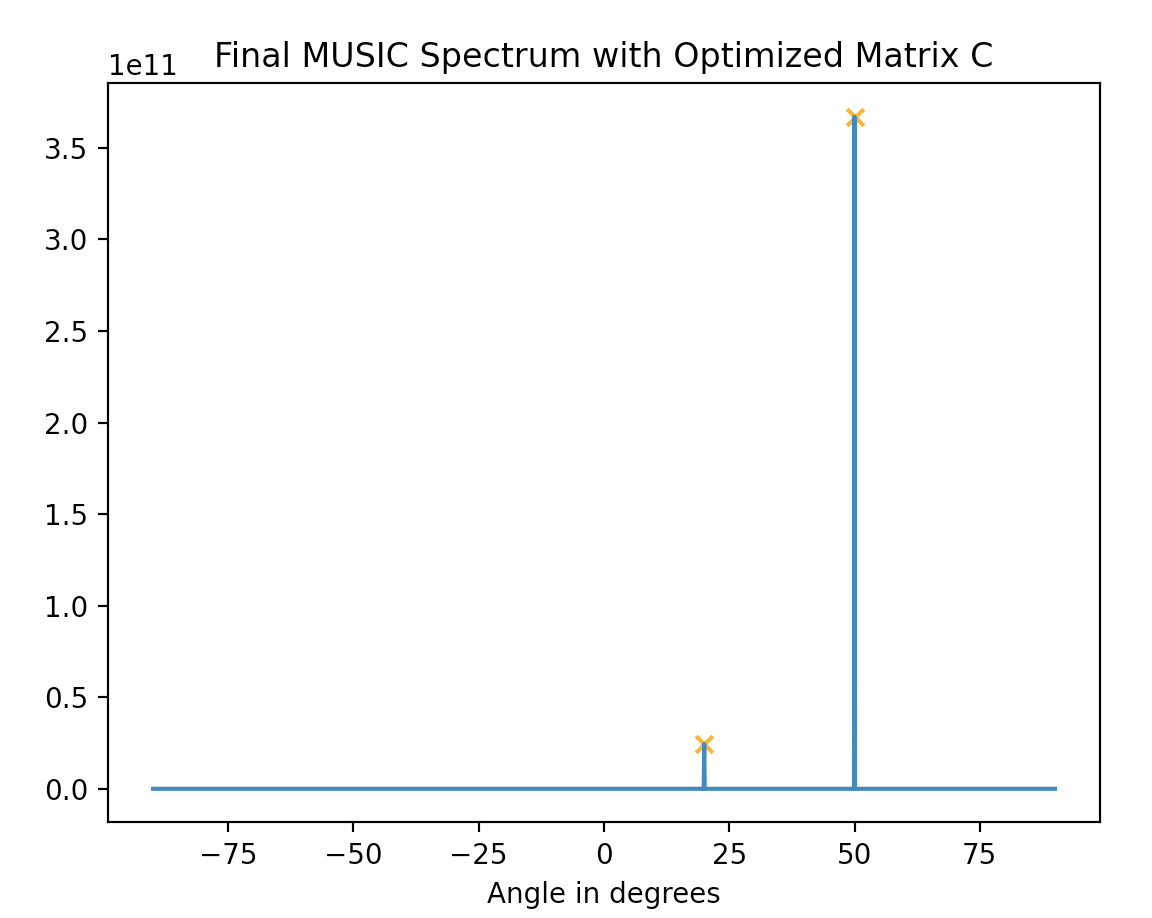}
        \caption{MUSIC spectrum for $\beta = 0.2$}
        \label{fig:0.2_spect}
    \end{subfigure}
     \begin{subfigure}{.18\linewidth}
        \centering
        \includegraphics[width = 0.95 \linewidth]{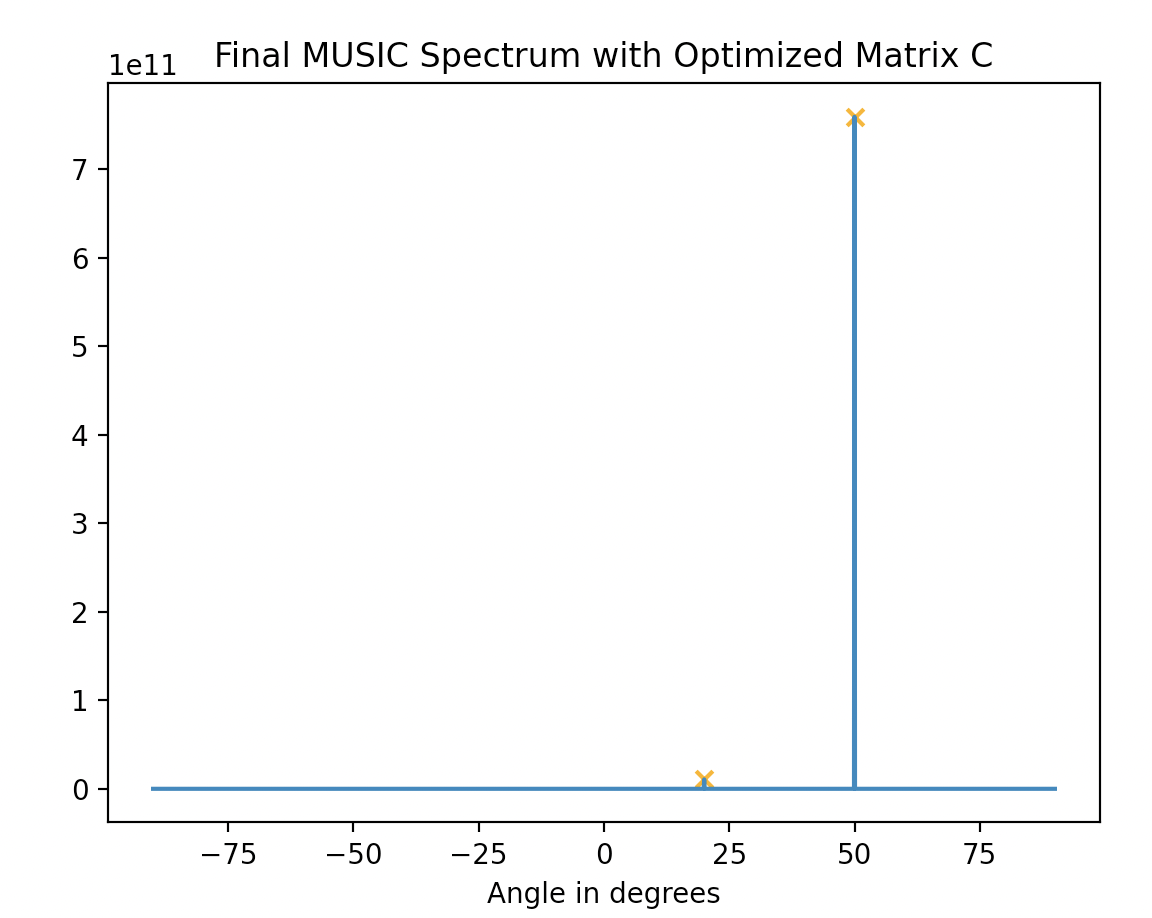}
        \caption{MUSIC spectrum for $\beta = 0.5$}
        \label{fig:0.5_spect}
    \end{subfigure}
    \begin{subfigure}{.18\linewidth}
        \centering
        \includegraphics[width = 0.95 \linewidth]{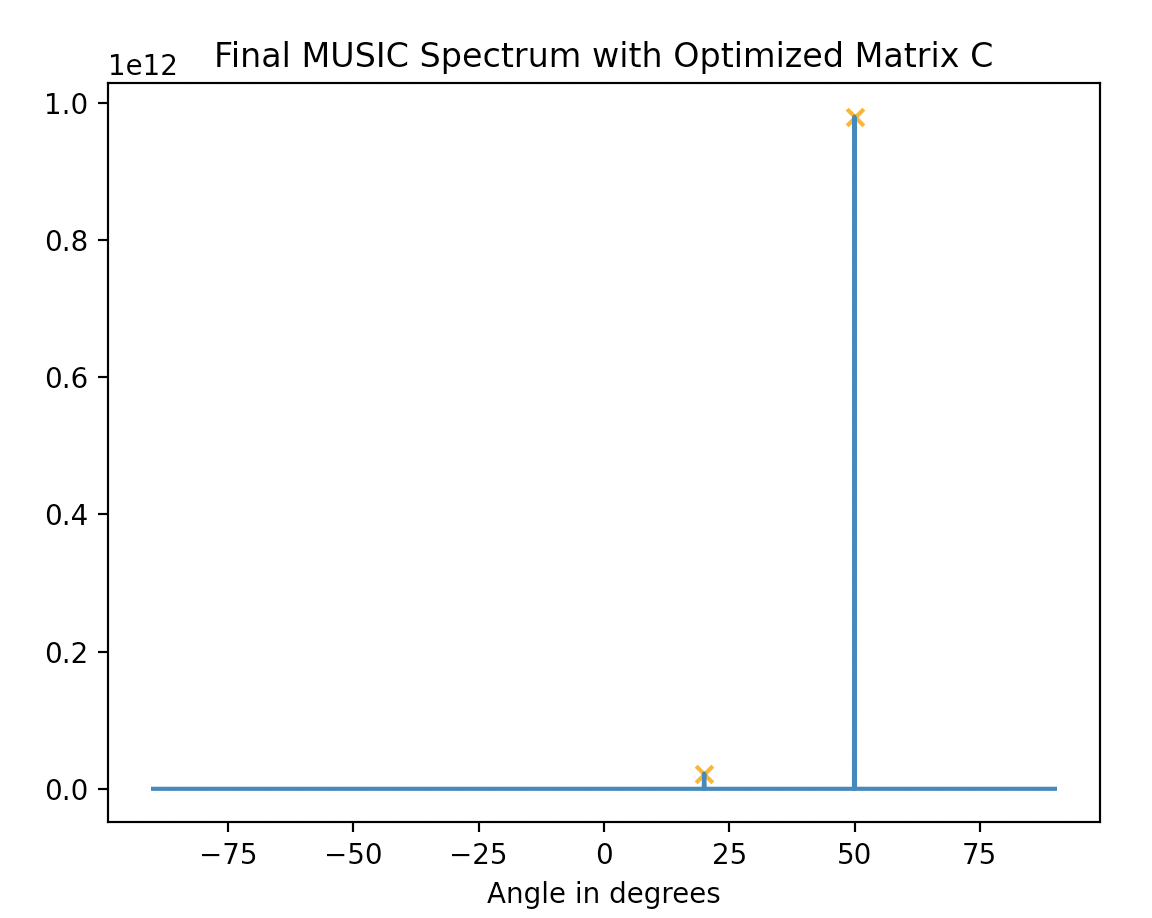}
        \caption{MUSIC spectrum for $\beta = 0.8$}
        \label{fig:0.8_spect}
    \end{subfigure}
    \begin{subfigure}{.18\linewidth}
        \centering
        \includegraphics[width = 0.95 \linewidth]{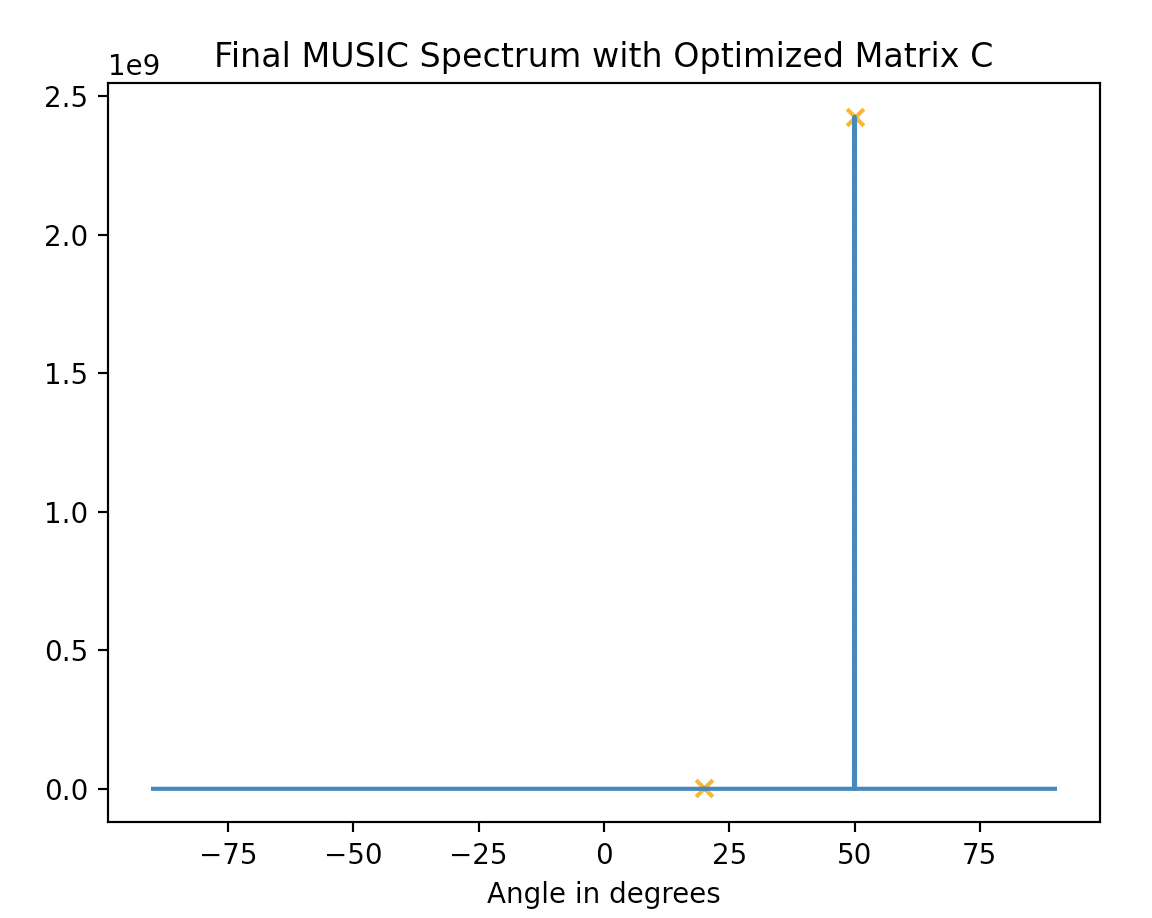}
        \caption{MUSIC spectrum for $\beta = 1$}
        \label{fig:1_spect}
    \end{subfigure} 
    \begin{subfigure}{.18\linewidth}
        \centering
        \includegraphics[width = 0.95 \linewidth]{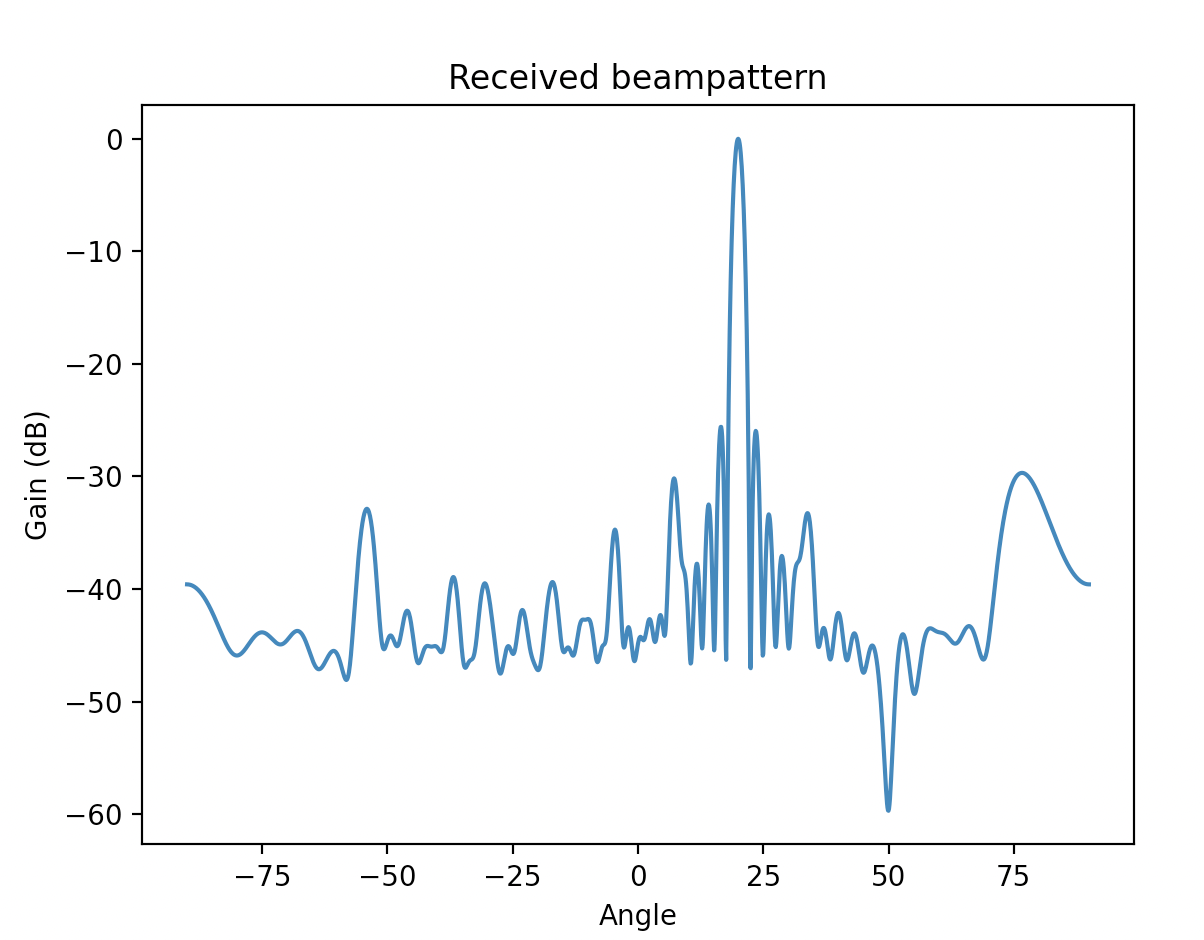}
        \caption{Beam pattern for \\ $\beta = 0$}
        \label{fig:0_pattern}
    \end{subfigure}
    \begin{subfigure}{.18\linewidth}
        \centering
        \includegraphics[width = 0.95 \linewidth]{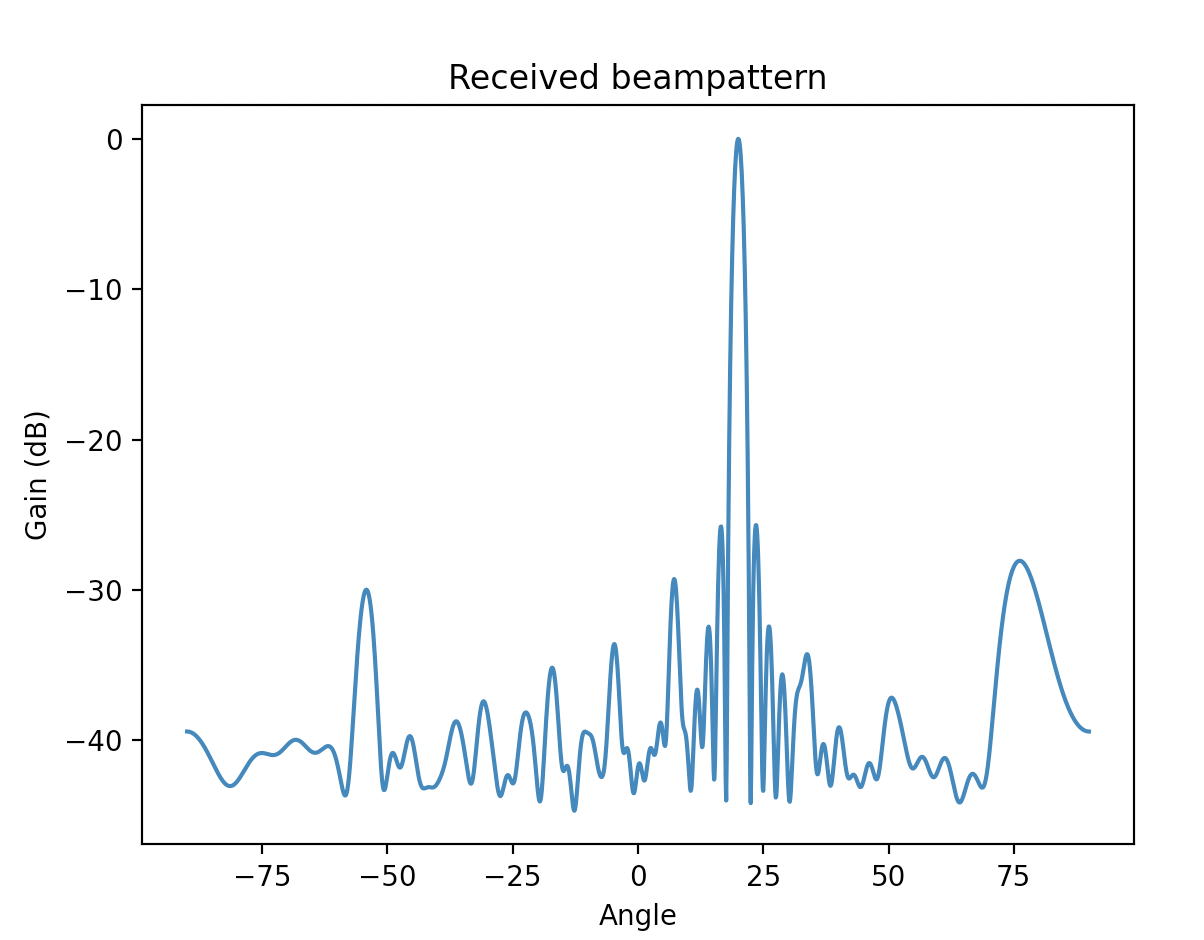}
        \caption{Beam pattern for \\ $\beta = 0.2$}
        \label{fig:0.2_pattern}
    \end{subfigure}
     \begin{subfigure}{.18\linewidth}
        \centering
        \includegraphics[width = 0.95 \linewidth]{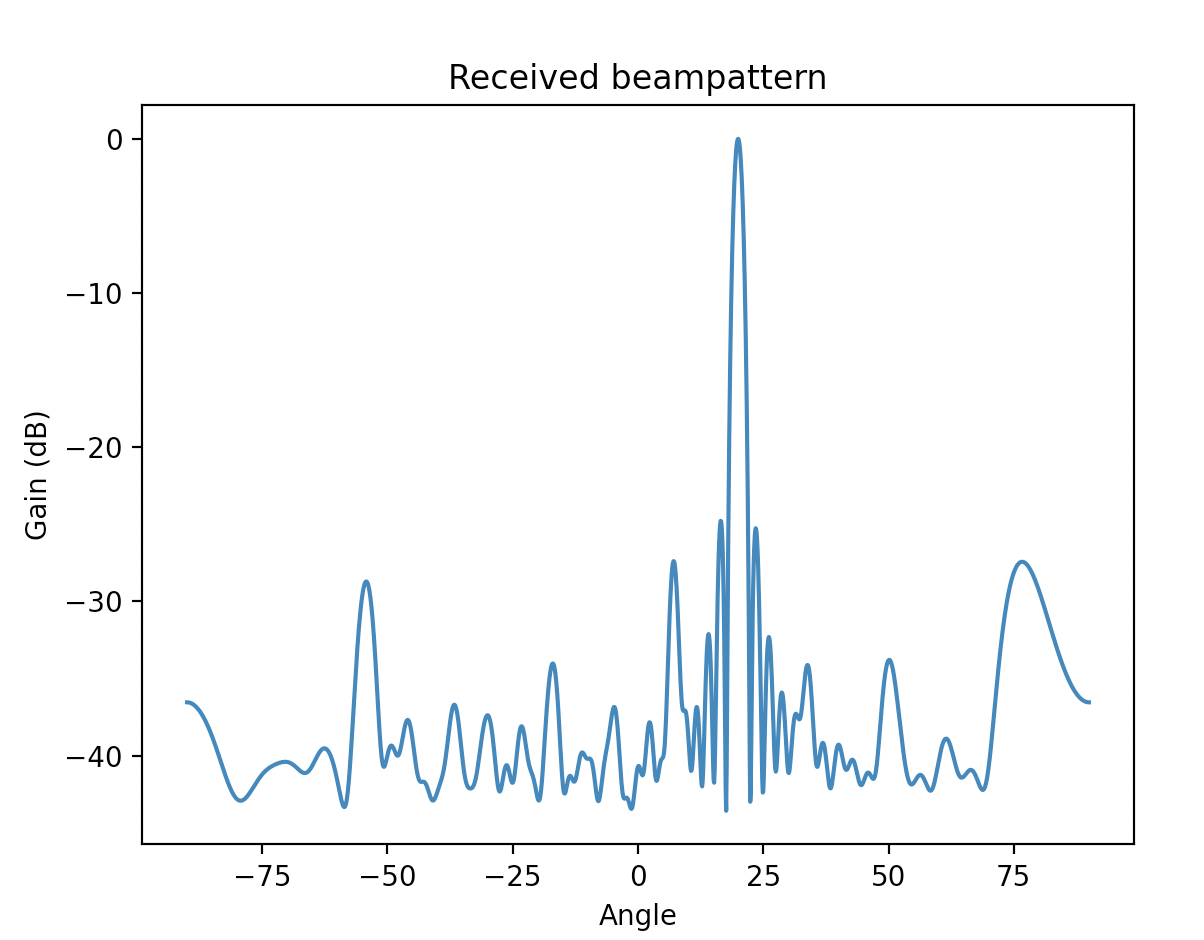}
        \caption{Beam pattern for \\ $\beta = 0.5$}
        \label{fig:0.5_pattern}
    \end{subfigure}
    \begin{subfigure}{.18\linewidth}
        \centering
        \includegraphics[width = 0.95 \linewidth]{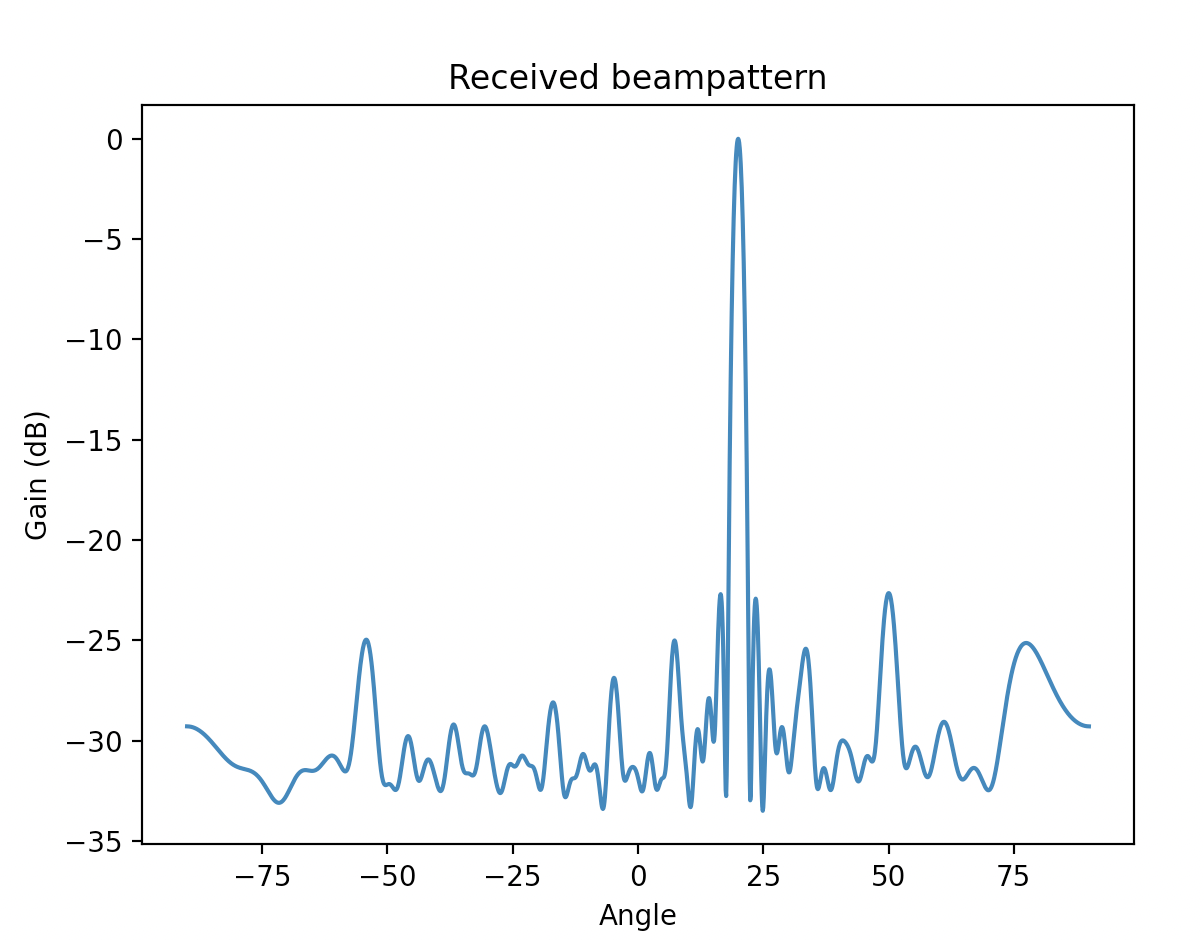}
        \caption{Beam pattern for \\ $\beta = 0.8$}
        \label{fig:0.8_pattern}
    \end{subfigure}
    \begin{subfigure}{.18\linewidth}
        \centering
        \includegraphics[width = 0.95 \linewidth]{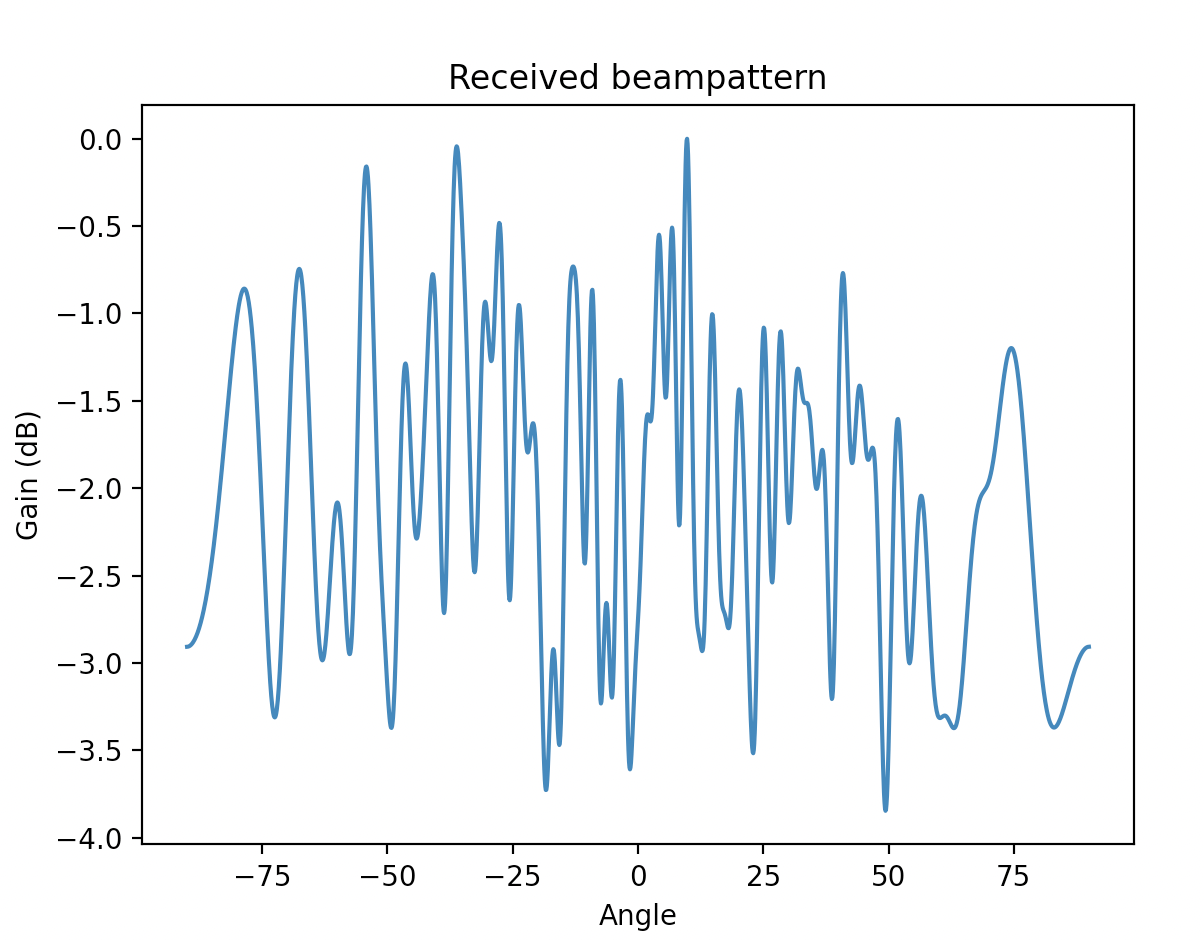}
        \caption{Beam pattern for \\ $\beta = 1$}
        \label{fig:1_pattern}
    \end{subfigure}  
    \begin{subfigure}{.18\linewidth}
        \centering
        \includegraphics[width = 0.95 \linewidth]{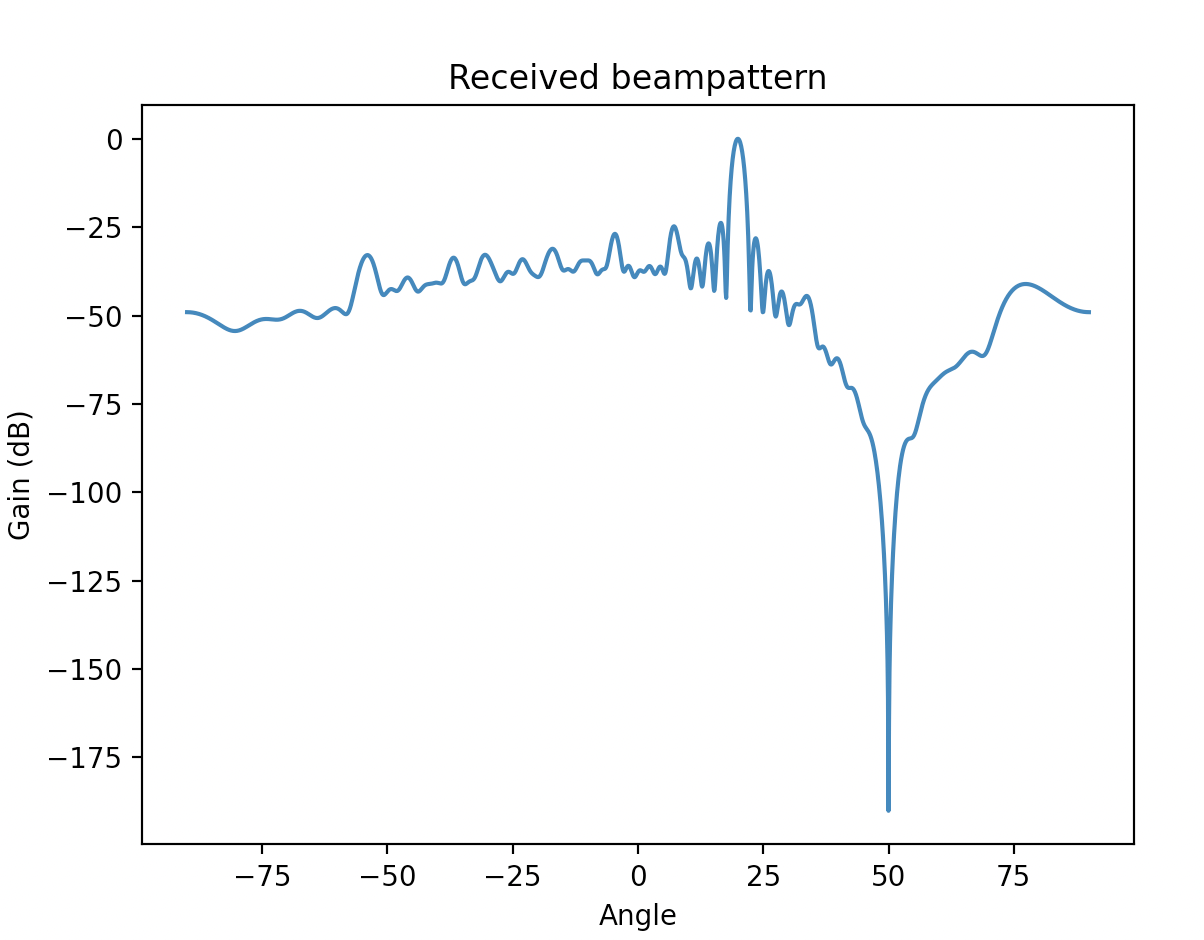}
        \caption{Beam pattern after convolution for $\beta = 0$}
        \label{fig:0_pattern_conv}
    \end{subfigure}
    \begin{subfigure}{.18\linewidth}
        \centering
        \includegraphics[width = 0.95 \linewidth]{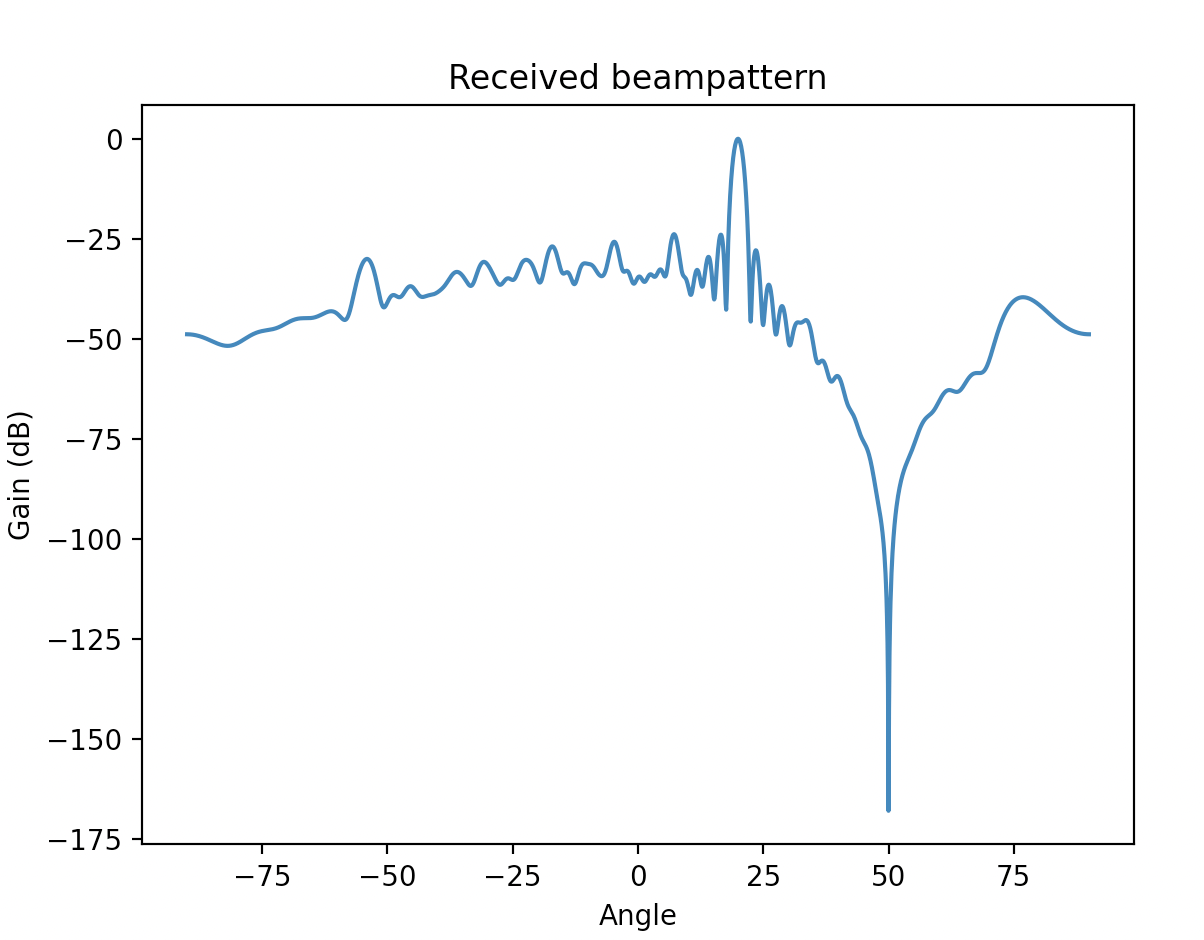}
        \caption{Beam pattern after convolution for $\beta = 0.2$}
        \label{fig:0.2_pattern_conv}
    \end{subfigure}
     \begin{subfigure}{.18\linewidth}
        \centering
        \includegraphics[width = 0.95 \linewidth]{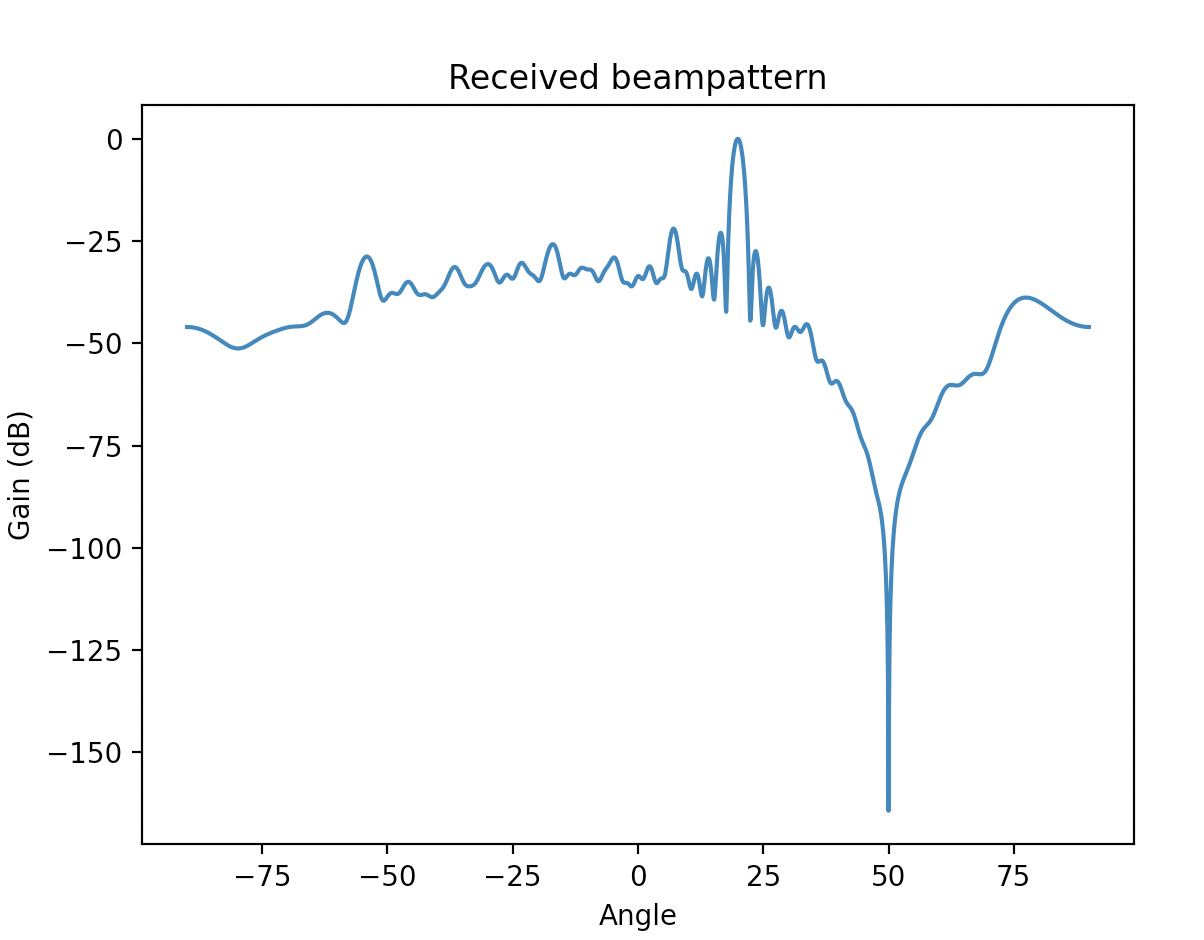}
        \caption{Beam pattern after convolution for $\beta = 0.5$}
        \label{fig:0.5_pattern}
    \end{subfigure}
    \begin{subfigure}{.18\linewidth}
        \centering
        \includegraphics[width = 0.95 \linewidth]{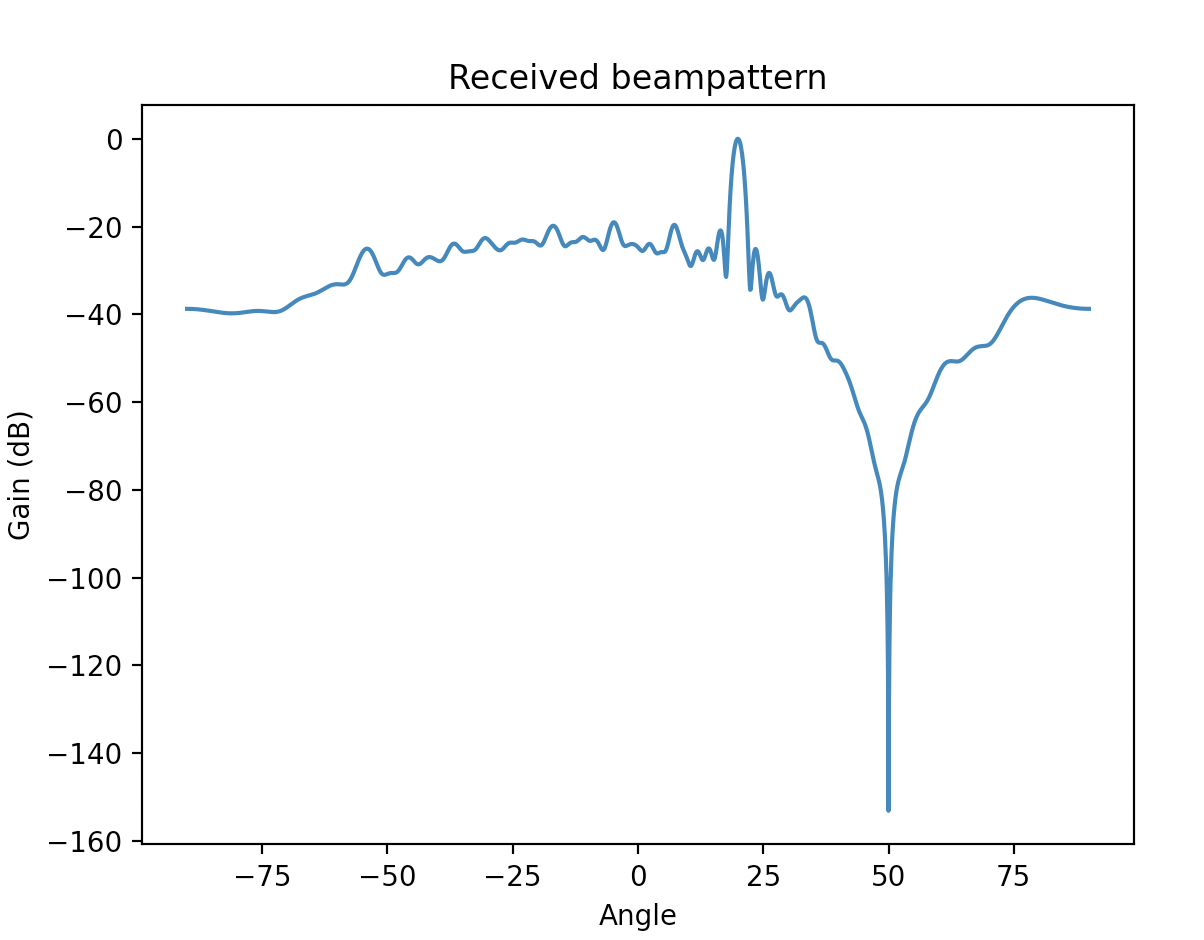}
        \caption{Beam pattern after convolution for $\beta = 0.8$}
        \label{fig:0.8_pattern}
    \end{subfigure}
    \begin{subfigure}{.18\linewidth}
        \centering
        \includegraphics[width = 0.95 \linewidth]{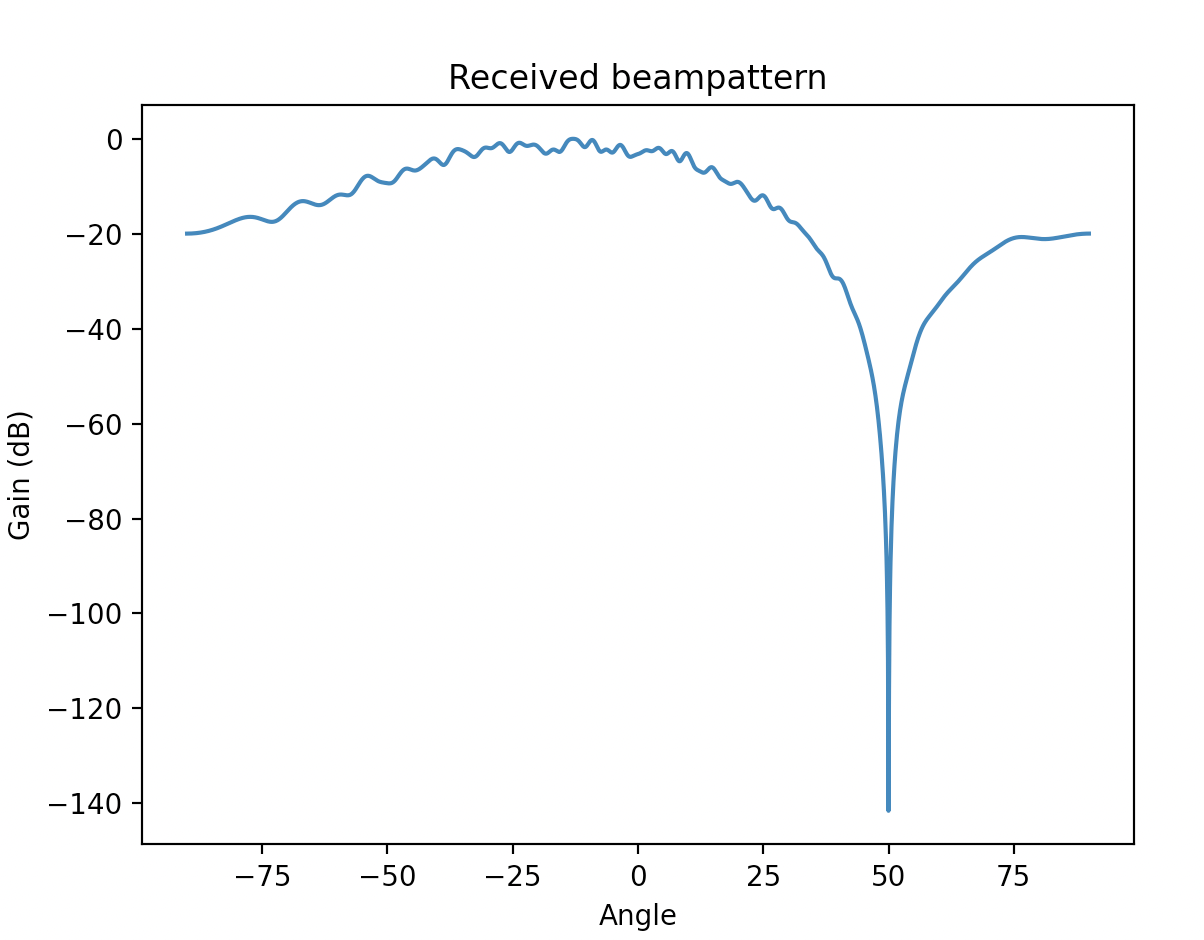}
        \caption{Beam pattern after convolution for $\beta = 1$}
        \label{fig:1_pattern}
    \end{subfigure}%
    \caption{Modified MUSIC spectrum (a) - (e), Beam pattern (f) - (j), and Beam pattern after convolution (k) - (o), for various $\beta$.}
    \label{fig:patterns}
\end{figure*}

The above loss ensures that both the target and the interference are detectable in the modified MUSIC algorithm. This may result in insufficient damping of the interference radar's power. To ensure stronger notch at the angle of the interference, one can find the new configuration of the RIS by convolving it with the term $[1, e^{-j \pi sin(\theta_i)}]$ \cite{convpaper}. This process will ensure sufficient notch at the angle of the interference. 

Until now, the calculations were performed over one subcarrier. The received signal consists of several subcarriers. As stated in Section \ref{sec:model}, the received signal can be decomposed to its subcarriers and symbols by applying the Fourier transform over the duration of the signal as:
\begin{equation}
    y[n,m] = \eta_{n} \boldsymbol{C_m}^T \boldsymbol{b}_{n}(\theta_t) + \eta_{i,n} \boldsymbol{C_m}^T \boldsymbol{b}_{n}(\theta_i)  + z[n,m],
\end{equation}
where $\eta_{n}$ and $\eta_{i,n}$ are the symbol and path gains of target and interference. Considering the stationary situation over the samples of each subcarrier, one can perform angle detection by using (\ref{eq:musiceq}) for each subcarrier. By averaging the estimated angles, one can create the input of the deep learning model. the new structure is displayed as Fig.~\ref{fig:structure} with the modified loss to account for all the subcarriers as:
\begin{align}
    L &= \beta \sum_{n = 0}^{N-1}\frac{\lVert \boldsymbol{Q}_{null}^H \boldsymbol{C}^T \boldsymbol{b}_n(\hat{\theta_i}) \rVert_2^2}{\lVert \boldsymbol{Q}_{null}^H \boldsymbol{C}^T \boldsymbol{b}_n(\hat{\theta_t}) \rVert_2^2} \\ \nonumber &+ (1-\beta ) \sum_{n = 0}^{N - 1} \frac{\lVert \boldsymbol{C}^T \boldsymbol{b}_n(\hat{\theta_i})\rVert_2^2 + \sigma^2}{ \lVert \boldsymbol{C}^T \boldsymbol{b}_n(\hat{\theta_t})\rVert_2^2}
\end{align}

\begin{figure}
    \centering
    \includegraphics[width=0.5\linewidth]{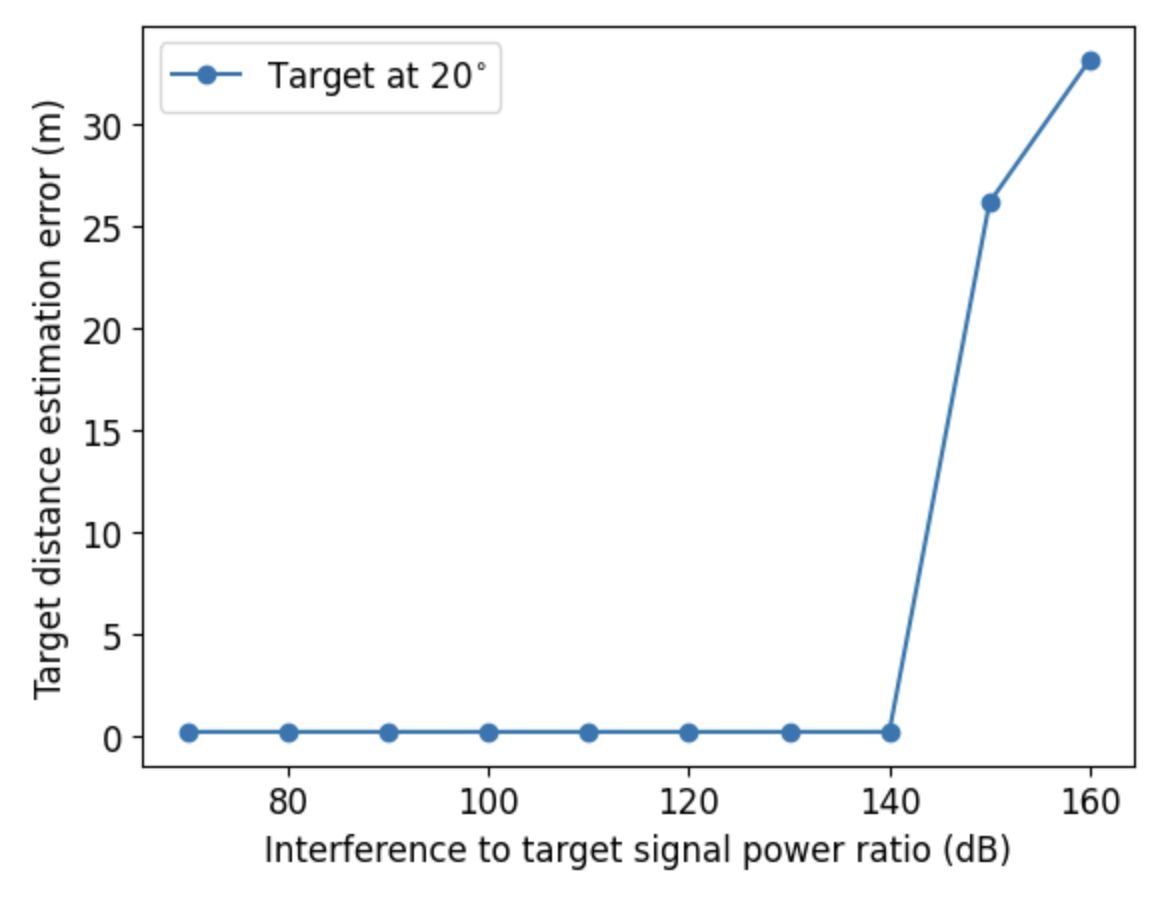}
    \caption{Error in estimation of the target based on the ratio of the received power from interference and target angle}
    \label{fig:target_error}
\end{figure}

\begin{figure*}
\centering
    \begin{subfigure}{.23\linewidth}
        \centering
        \includegraphics[width = 0.95 \linewidth]{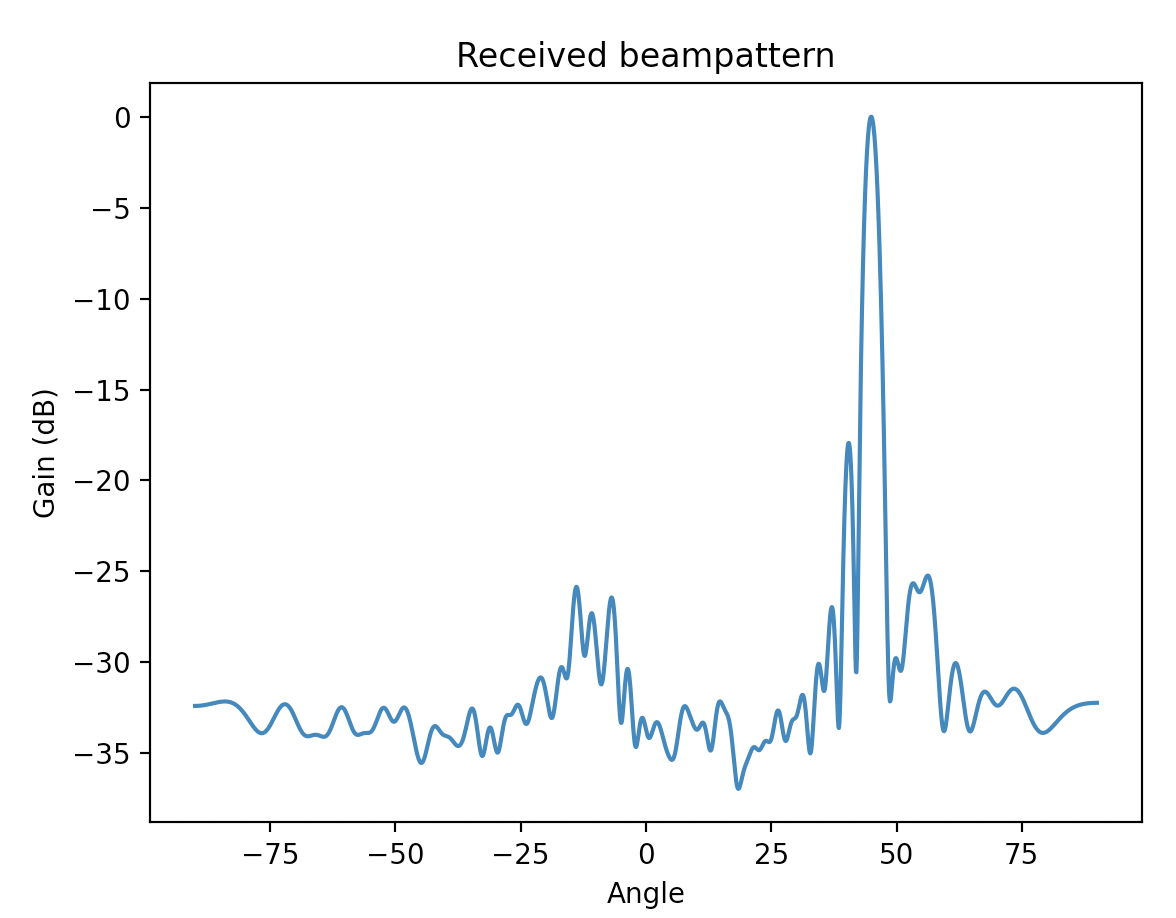}
        \caption{Beam pattern for target at $45^{\circ}$}
        \label{fig:45_raw}
    \end{subfigure}
    \begin{subfigure}{.23\linewidth}
        \centering
        \includegraphics[width = 0.95 \linewidth]{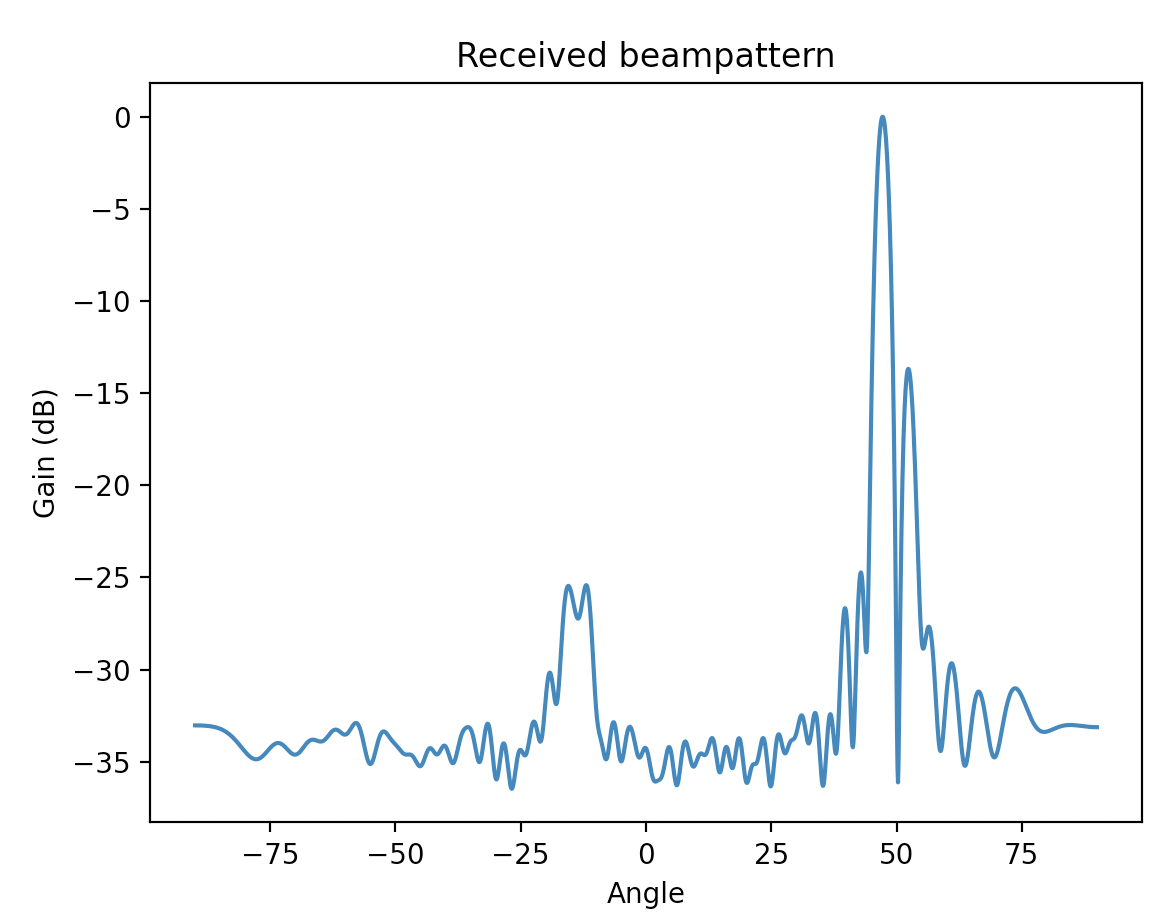}
        \caption{Beam pattern for target at $48^{\circ}$}
        \label{fig:48_raw}
    \end{subfigure} 
     \begin{subfigure}{.23\linewidth}
        \centering
        \includegraphics[width = 0.95 \linewidth]{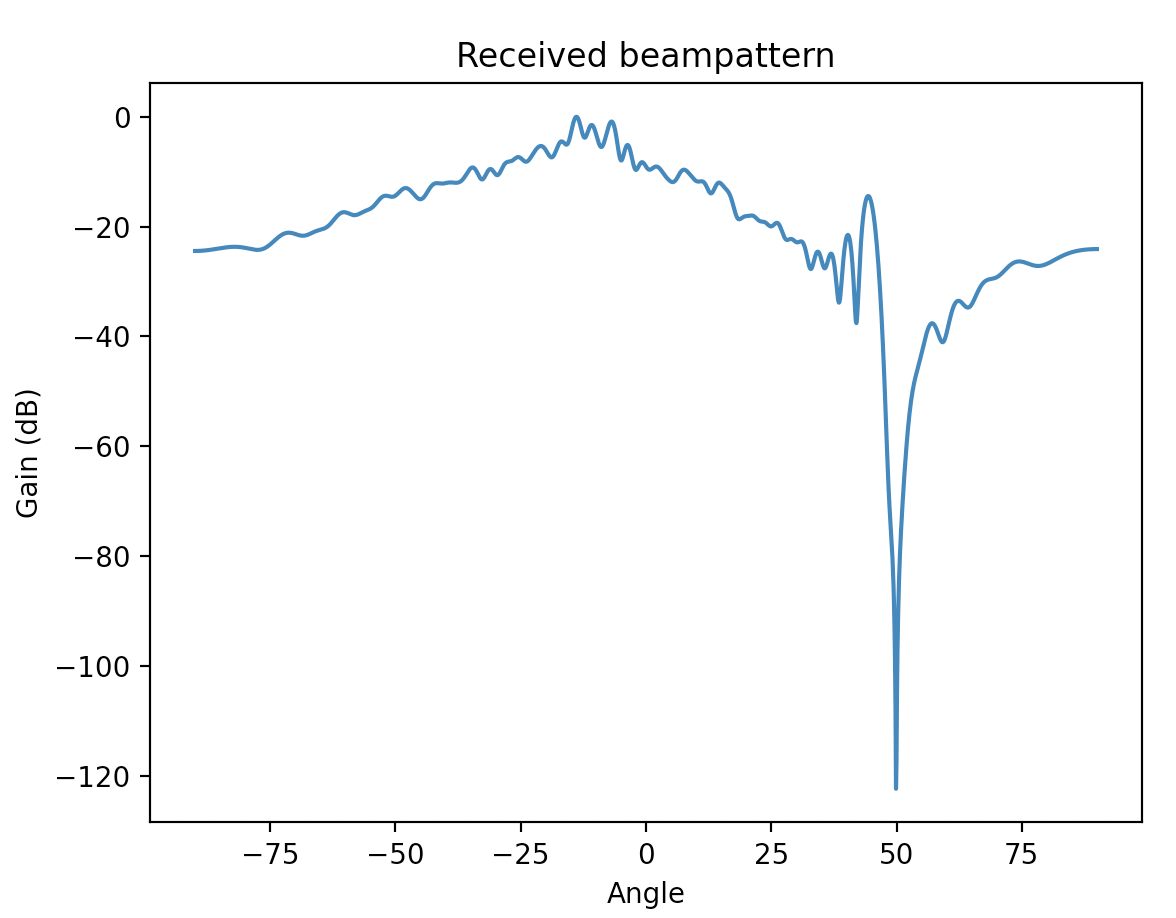}
        \caption{Beam pattern after convolution for target at $45^{\circ}$}
        \label{fig:45_conv}
    \end{subfigure}
    \begin{subfigure}{.23\linewidth}
        \centering
        \includegraphics[width = 0.95 \linewidth]{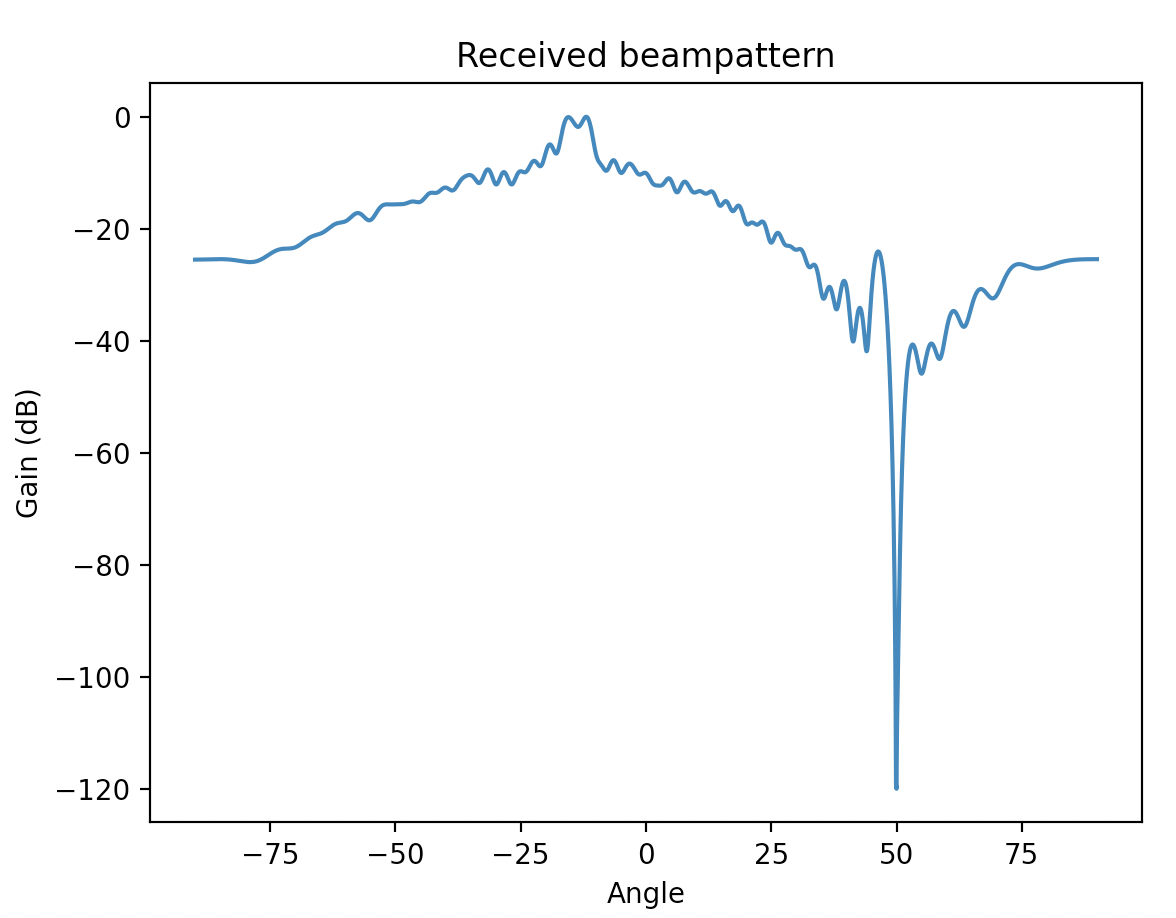}
        \caption{Beam pattern after convolution for target at $48^{\circ}$}
        \label{fig:48_conv}
    \end{subfigure}%
    \caption{Beam pattern for closely spaced target and interference.}
    \label{}
\end{figure*}

\section{Simulation Results} \label{sec:simulation}

In this section, the simulation results are presented. Consider the victim radar operating at $f_c = 77 GHz$, with a bandwidth of $B = 200 MHz$. The radar operates with $N = 20$ subcarriers and $M = 100$ samples per each subcarrier with the samples selected from a PSK modulation scheme. A RIS is located at the center of the coordinate system with $L = 50$ elements. 

Consider the target at the unknown angle of $\theta_t = 20^{\circ}$ with a relative distance of $R= 30m$ and the interference radar with an unknown angle of $\theta_i = 50^{\circ}$ with relative distance of $R_i = 15m$. The interference radar has the same characteristics as the victim radar. Let us first explore the created pattern by the algorithm. 

Let us consider the structure of the deep learning model. One can use two linear layers with the ReLU activation function as the middle layers and the last layer as a linear layer. Fig.~\ref{fig:patterns}, shows the created beam pattern as well as the MUSIC spectrum of the first subcarrier for various values of $\beta$. It can be seen that by increasing $\beta$, the spectrum will have stronger peak at the angle of the interference, which ensures better detection and more accurate estimate of the angle. Despite better estimation, the relative gain of the target's angle with respect to the interference angle at the beam pattern becomes smaller, to the extent of no gain in the case of $\beta = 1$. It can be seen that the pattern after the convolution further decreases the notch and increases the ratio of the gain of the target and the interference angles. Decreasing $\beta$ will increase the gain of the target compared to the interference angle, but it will also make the modified MUSIC's spectrum peaks near each other, which has the potential of picking the interference angle instead of the target angle, resulting in wrong interpretation. It is observed that mid-range to upper range values of $\beta$, e.g. $\beta = 0.8$ produce a good balance of the pattern as well as the spectrum peak. Furthermore, the pattern can be reinforced using the convolution technique \cite{convpaper}, making the mid-range to upper range $\beta$ a viable option.

Next we explore the effectiveness of the algorithm in terms of detecting the target range. Fig.~\ref{fig:target_error} shows the error in estimation of the location of the targets based on the maximum likelihood estimation (MLE) technique as outlined in \cite{rvmap}. It can be seen that the method performs well in removing the effect of the interference, but after some point where the power of the interference exceeds the RIS gain, the error increases.

We also investigate how sensitive the approach is in terms of detecting the angles. We assume that the target and the interference are relatively close to each other. Fig.~\ref{fig:45_raw} shows the beam pattern for the case of a target at $45^{\circ}$ and the interference at $50^{\circ}$. It can be seen that the algorithm can provide a pattern to enhance the gain of the target and reduce the power of the interference. Also, this can be further improved by using the convolution as depicted in the figure. Additionally, the algorithm performs well when the target and the interference are $2^{\circ}$ apart. Fig.~\ref{fig:48_raw} shows the case of a target at $48^{\circ}$ and the interference at $50^{\circ}$. It can be seen that the model creates a peak at 48 degrees and a notch at $50$ degrees. By applying the convolution, the notch goes further down, and the peak is reduced, however the overall ratio of the target gain to interference gain improves. Furthermore, the error in the estimation of the angles in all cases were less than $0.001^{\circ}$.

\section{Conclusion} \label{sec:conclusion}
In this paper, we proposed a deep learning-based framework for enhancing OFDM radar systems affected by interference, through the use of RIS. The proposed method effectively estimates the angles of the target and interference while optimizing the RIS configuration to minimize the impact of interference. Our deep learning model, driven by a carefully designed loss function, ensures that both the target and interference peaks are detectable in the modified MUSIC algorithm. The approach is further extended to handle multiple subcarriers, improving its applicability to real-world scenarios. Simulation results demonstrate the effectiveness of the method in increasing the signal-to-interference-plus-noise ratio (SINR) and improving the target localization accuracy. The proposed technique offers a solution for enhancing radar performance in challenging environments with interference while maintaining robust and accurate angle estimates.



\ifCLASSOPTIONcaptionsoff
  \newpage
\fi



%



\bibliographystyle{IEEEtran}
\bibliography{bibtex/bib/IEEEexample}

%








\end{document}